\documentclass[pdflatex,sn-mathphys]{sn-jnl}% Math and Physical Sciences Reference Style
\jyear{2021}%

\theoremstyle{thmstyleone}%
\theoremstyle{thmstyletwo}%

\theoremstyle{thmstylethree}%
\usepackage{comment}

\usepackage{lineno}
\begin{document}

\title[%Interacting internal waves explain global patterns of interior ocean mixing
]{Interacting internal waves explain global patterns of interior ocean mixing}


\author*[1,2]{Giovanni Dematteis}%\email{giovannidematteis@gmail.com}
\author[3]{Arnaud Le Boyer}

\author[4]{Friederike Pollmann}
\author[2]{Kurt L. Polzin}
\author[3]{Matthew H. Alford}
\author[5]{Caitlin B. Whalen}
\author[6]{Yuri V. Lvov}

\affil[1]{{Universit\`a degli Studi di Torino}, {{Torino}, {Italy}}}
\affil[2]{{Woods Hole Oceanographic Institution}, {{Woods Hole}, {MA}, {USA}}}
\affil[3]{{Scripps Institution of Oceanography}, {{La Jolla}, {CA}, {USA}}}
\affil[4]{{Institut für Meereskunde}, {Universit\"at Hamburg}, {{Hamburg}, {Germany}}}
\affil[5]{{Applied Physics Laboratory}, {University of Washington}, {{Seattle}, {WA}, {USA}}}
\affil[6]{{Department of Mathematical Sciences}, {Rensselaer Polytechnic Institute}, {{Troy}, {NY}, {USA}}}

\affil[*]{giovannidematteis@gmail.com}

\abstract{Across the stable density stratification of the abyssal ocean, deep dense water is slowly propelled upward by sustained, though irregular, turbulent mixing.
The resulting mean upwelling determines large-scale oceanic circulation properties like heat and carbon transport. 
In the ocean interior, this turbulent mixing is caused mainly by breaking internal waves: generated predominantly by winds and tides, these waves interact nonlinearly, transferring energy downscale, and finally become unstable, break and mix the water column.
This paradigm, long parameterized heuristically, still lacks full theoretical explanation.
Here, we close this gap using wave-wave interaction theory with input from both localized and global observations.
We find near-ubiquitous agreement between first-principle predictions and observed mixing patterns in the global ocean interior. 
Our findings lay the foundations for a wave-driven mixing parameterization for ocean general circulation models that is entirely physics-based, which is key to reliably represent future climate states that could differ substantially from today's.}




\maketitle

\section*{Introduction}\label{sec1}

% Introduction currently structured in 5 paragraphs as suggested by Caitlin

%{\it Paragraph 1: Internal wave driven mixing is important globally} 

Turbulent vertical mixing in the ocean occurs when intermittent turbulent eddies smaller than a few meters, so-called oceanic microstructures, are generated in stratified regions through shear or convective instabilities. In the ocean's interior, these instabilities are primarily caused by the breaking of internal waves -- energetic oscillations free to propagate through the density-stratified ocean bulk. %, with 
Providing a mechanism for dense water to slowly rise from the deep ocean \cite{Ferrari,garabato2022oceanBook}, internal wave-driven vertical mixing is generally believed to be one of the main drivers of the oceanic circulation \cite{garrett1979internal,garabato2022oceanBook}. It thus shapes the Earth's climate~\cite{melet2022role,de2022role,cimoli2023significance}, influencing, among others, sea level rise~\cite{hallberg2013sensitivity}, nutrient fluxes and hence marine ecosystems \cite{bindoff2019changing}, and anthropogenic heat and carbon uptake \cite{song2019impact}.

%{\it Paragraph 2: To aid in global observations and modeling it is key that we have simple ways of understanding the link between internal waves and mixing}

The small scales (below a few meters) and intermittent nature (timescales of minutes to hours) of turbulent mixing imply that it is both difficult to observe directly and impossible to resolve in ocean general circulation models (OGCM), whose grid cells are substantially larger than the dominant turbulence length scales. Methods that allow us to infer wave-induced mixing from far more easily obtainable observations at larger scales (from a few to hundreds of meters) -- the so-called finestructure -- are hence central to advancing our understanding of ocean mixing processes and how to best parameterize them in numerical models~\cite{whalen2020internal,garabato2022oceanBook}.

%{\it Paragraph 3: Our current best simple way is (describe way) which is commonly called the finescale parameterization }

The state-of-the-art method to infer internal wave-driven mixing is to interpret turbulent mixing as the energy sink at the end of a downscale energy cascade through the oceanic internal wavefield, fueled by large-scale forcing and sustained by wave-wave interaction processes~\cite{olbers1976nonlinear,Muller86,polzin2014finescale,mackinnon2017climate}. This picture {\color{black}was initially supported by formal theory~\cite{olbers1976nonlinear,Muller86} and ray-tracing numerical simulations~\cite{henyey1986energy}} and then captured {\color{black}heuristically} by the Finescale Parameterization (FP) formula~\cite{gregg1989scaling,henyey1991scaling,wijesekera1993application,polzin1995finescale}, essentially in the following form:

\begin{equation}\label{eq:0}
    \epsilon = \epsilon_0 \frac{f}{f_0} \frac{N^2}{N_0^2}\frac{{\mathcal{E}}^2}{{\mathcal{E}}_0^2}\,,
\end{equation}
allowing for an estimation of mixing metrics -- here, the turbulent kinetic energy (TKE) dissipation rate $\epsilon$ associated with the decay time of the small-scale turbulent eddies -- from characteristics of the internal wavefield: its shear-variance level ${\mathcal{E}}$ (related to the energy level $E$), the local Coriolis frequency $f$, and the buoyancy frequency $N$. The subscript $0$ denotes reference values. {\color{black}Phenomenological corrections due to different spectral shapes \citep{polzin1995finescale,polzin2014finescale,ijichi2015frequency} are omitted in \eqref{eq:0} for simplicity. We refer to \eqref{eq:0} as the Gregg-Henyey-Polzin (GHP) parameterization ~\cite{polzin2014finescale}, which requires input from both shear and strain spectra. Strain-based~\cite{wijesekera1993application} and shear-based~\cite{gregg1989scaling} versions exist -- below, we will compare our results with estimates from strain-based FP.} 
Although the internal wave-driven mixing picture is known to break down in particular locations of the ocean,
\textcolor{black}{notably near the ocean's surface and bottom boundaries and in regions with strong currents, high mesoscale activity ({\color{black} where interactions between geostrophically balanced motions and internal waves are large}), and the presence of fronts}~\citep{watermanSuppressionInternalWave2014,klymakDirectBreakingInternal2008,
cusack2020observed,garabato2022kinetic,whalen2018large,danioux2015concentration,chouksey2022gravity}, the FP framework is corroborated by substantial observational evidence around the global ocean~\cite{polzin1995finescale,whalen2012spatial,waterhouse2014global, pollmann2023, kunze2017internal,pollmann2020global,le2021variability} and is also supported by {\color{black}recent} numerical evaluations of the scattering integral of wave-wave interactions \cite{eden2019numerical,dematteis2022origins}. The FP approach has made it possible to estimate global maps of $\epsilon$~\cite{whalen2012spatial,waterhouse2014global,pollmann2023} in the interior ocean, and to parameterize mixing in an energetically constrained way \cite{olbers2019idemix, melet2013sensitivity} 
in some of the newest versions of the OGCMs.

%{\it Paragraph 4: This way has X, Y, and Z conceptual/theoretical gaps}
However, significant conceptual research gaps persist in the FP framework: (i) The constant $\epsilon_0$ in Eq.~\eqref{eq:0} is empirical~\cite{polzin2014finescale}, but in order to reliably represent a wide range of conditions in a changing climate~\cite{zhang2021decreased} the parameterized link between internal waves and ocean mixing needs to be based on the underlying physics~\cite{whalen2020internal}; (ii) The formula centers around the 1976 version of the Garrett-Munk spectrum~\cite{GM76} (GM76) accounting for departures from it via phenomenological correction -- omitted in~\eqref{eq:0} -- but new observational knowledge that GM76 is one out of many realistic spectra~\cite{alford2007seasonal,regional,pollmann2020global,le2021variability} calls for a process-based formula that applies to any spectral shape indistinctly;  (iii) The interpretation of~\eqref{eq:0} relies on the notion that wave-wave interactions with large scale separation dominate the energy transfers~\cite{Muller86}, 
but this paradigm requires a nonlinearity level arguably too strong for the assumed weakly nonlinear theory to apply~\cite{holloway1980oceanic} and the presence of a high-frequency source not backed-up by observations~\cite{polzin2014finescale,mackinnon2017climate}.

%{\it Paragraph 5: Here we fill these gaps for the first time and show that the improved version matches common observations of epsilon}
Here, {\color{black}we pursue the application of formal theory~\cite{olbers1976nonlinear,LT2} in a way that is more flexible to address the observed variability of the oceanic internal wavefield.} {\color{black}We show that the bare theory of wave-wave interactions (weak wave turbulence)~\cite{VLvov1992,NazBook,galtier_physics_2022} provides a systematic explanation for} the close causal link between the observed global patterns of internal wave spectral energy and of turbulent vertical mixing in the ocean interior. 
{\color{black}A few comments are due on the difference from important earlier attempts to trace the FP estimates back to first principles, in the ray-tracing %\textcolor{brown}{not defined/referenced yet}
regime of scale separation between small-amplitude small-scale test waves and a large-scale background shear~\cite{henyey1986energy,sun1999internal,ijichi2017eikonal}: (i) The ray-tracing simulations are implemented in the Wentzel-Kramers-Brillouin (WKB) approximation. This requires an arbitrary scale-separation factor that is used to tune the magnitude of the estimate, whereas the weak wave turbulence framework does not assume scale separation and allows for a quantification of spectrally local transfers; (ii) The spectrally-local transports of weak wave turbulence make the results rather independent of the high-wavenumber cutoff ({\it Supplementary Information}), {vs} a high sensitivity to the cutoff scale in the eikonal nonlocal transports~\cite{ijichi2017eikonal}; (iii) Both are weakly nonlinear theories and suffer from high levels of nonlinearity as the high-wavenumber cutoff is approached, but our analysis ({\it Supplementary Information}) shows that nonlinearity is not as high as Holloway's objection~\cite{holloway1980oceanic} depicted it. An updated rigorous analysis on the eikonal approach is found in~\cite{lvov2024generalized,polzin2022one}.}

We synthesize our findings into an adaptive parameterization that generalizes the FP formula and provides an expression that %, for the first time,
is based {\color{black} unambiguously (without tunable parameters)} on the ocean's primitive equations themselves instead of empirically established numbers. Crucially, we depart from the concept of a universal spectrum and fully encompass the observed variability of oceanic internal wave spectra, including an extra degree of freedom for the near-inertial spectral content, {\color{black}which has not yet been included in a theoretical framework before beyond a brief mention in \cite{polzin1995finescale}}.  Moreover, the inter-scale energy transfers that we quantify and characterize turn out to be dominated by spectrally-local interactions rather than scale-separated ones, to require large-scale energy sources consistent with observations, and to occur in a weakly nonlinear regime for most of the internal wave scales.
\textcolor{black}{We then corroborate our results with data from some of the most advanced oceanographic field programs, finding near-ubiquitous agreement between our theoretical predictions, input from direct microstructure observations, and other pre-existing estimates from FP methods}.

\begin{figure*}[ht]%
\centering
\includegraphics[width=1\textwidth]{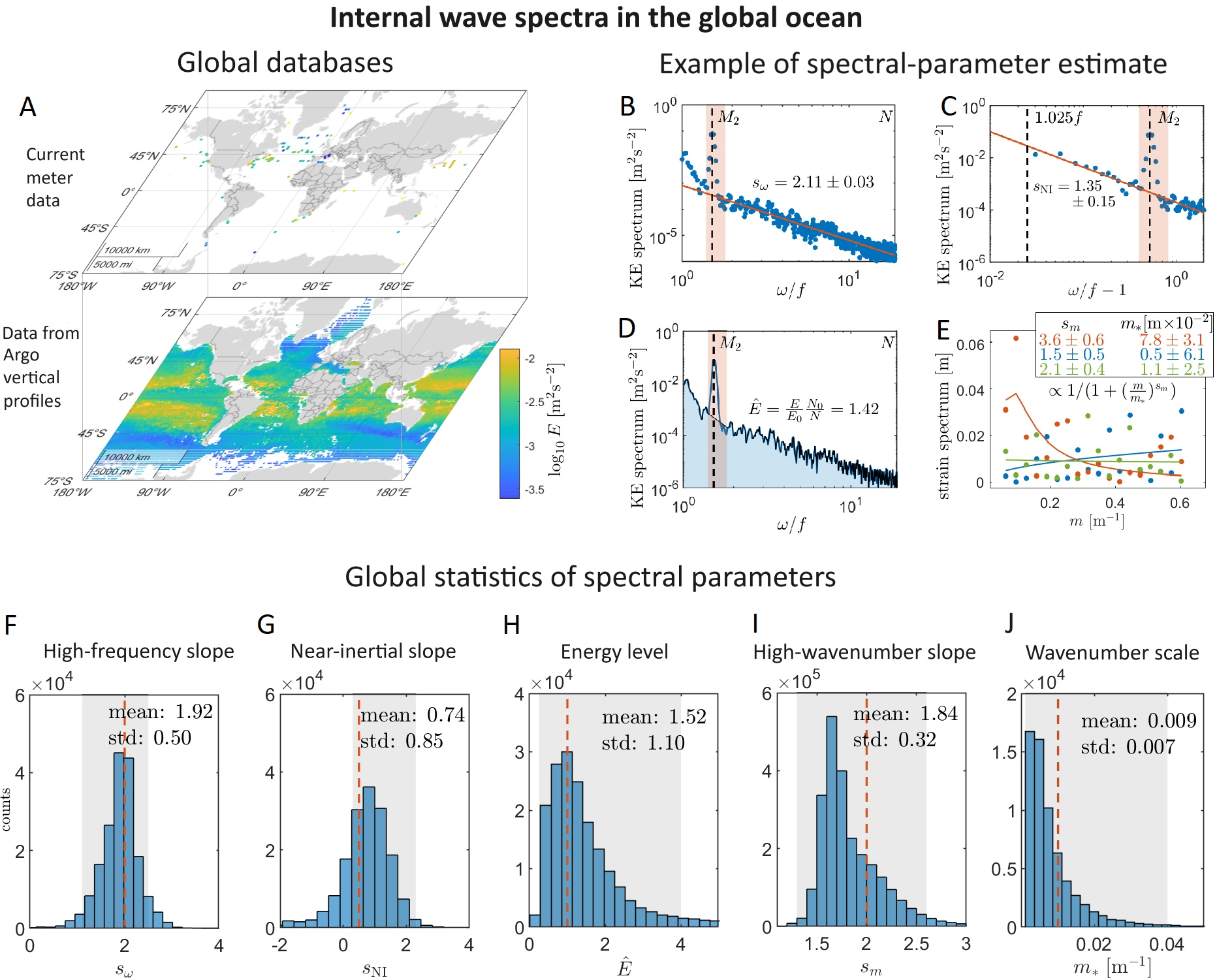}
\caption{{\bf Observed internal wave spectra and their global variability.} Input data for our global data set of two-dimensional spectral estimates of internal wave energy $e(m,\omega)$. Panel {\bf A} illustrates data availability of total internal wave energy in the two databases used {\color{black} ($250-500$ m depth range shown)}, the frequency spectra from current meter observations of  Global Multi-Archive Current-Meter Database (GMACMD, upper) and the vertical wavenumber spectral estimates from Conductivity-Temperature-Depth (CTD)-profiles collected by Argo floats (lower). The spectral parameters required for our theoretical model are obtained from fitting power-laws to, and integrating, the observed frequency spectra (the {\color{black}negative }spectral ``slopes'' at high frequency, $s_{\omega}$ in panel {\bf B}, and low frequency, $s_{\rm NI}$ in panel {\bf C}, and the normalized total energy $\hat{E}$ in panel {\bf D}; {\color{black}notice that the $M_2$ tidal peak is filtered out in our parameter estimation procedure}), and by nonlinear curve-fitting to observed wavenumber spectra to obtain the wavenumber {\color{black}negative} slope $s_m$ and scale $m_\star$ (panel {\bf E}). {\color{black}These exemplary spectra are from measurements at: $39.5^\circ$N, $54.1^\circ$W, $1006$ m depth ({\bf B}-{\bf D}); $29.6^\circ$S, $2.8^\circ$E, $589$ m depth ({\bf E, red}); $41.6^\circ$S, $46.8^\circ$W, $589$ m depth ({\bf E, blue}); $4.6^\circ$N, $29.1^\circ$W, $566$ depth ({\bf E, green})}. Panels {\bf F}-{\bf J} show the distribution of these spectral estimates for locations where all five of them are available. The {\color{black}grey-shaded backgrounds} represent the ranges of values covered by our implementation of the theoretical formula~\eqref{eq:2a}, {\color{black}while the red dashed lines identify the canonical Garrett-Munk spectrum (GM76) parameter values.}}
\label{fig:1}
\end{figure*}

\section*{Results}

\subsection*{Global variability of internal wave spectra}
\label{sec2}

We describe the spatially and temporally \textcolor{black}{observed oceanic variance as an internal wavefield -- purposely discard any vortical modes influence on such variance --  and characterize this internal wavefield in the two-dimensional (2D) frequency ($\omega$) and magnitude of vertical wavenumber ($m$) spectral space}. Internal waves oscillate with frequencies between $f$ and $N$, and vertically between the first mode, $m_0$ (inverse ocean depth), and the largest available mode that is not subject to shear instability~\cite{garabato2022oceanBook}, $m_c$ (see Eq.~\eqref{eq:mc}).

The energy spectrum, $e(m,\omega)$, represents the averaged distribution of energy in the internal-wave band, 
so that its 2D integration over this rectangular spectral region
amounts to the total internal-wave energy {\color{black}$E=KE+APE$, sum of kinetic and available potential energy. We neglect the contribution of vertical kinetic energy, thereby identifying $KE$ with the contribution from the horizontal velocity only}. To capture the variability of the internal wavefield in the ocean, we propose to represent the energy spectrum by five parameters ({\it Methods}~\ref{sec:6.2.1}, Eq.~\eqref{eq:en-spec}): the high-frequency and high-wavenumber {\color{black}negative slopes}, $s_\omega$ and $s_m$; the finite-point singularity exponent as $\omega\to f$, $s_{\rm NI}$, encoding the concentration of energy in the so-called inertial peak; the wavenumber scale under which the low-wavenumber energy density saturates, $m_*$; and the total energy $E =\hat E E_0 N/N_0$, with $E_0=0.003 $ m$^2 $s$^{-2}$. The GM76 spectrum is recovered by setting $s_\omega=2$, $s_m=2$, $s_{\rm NI}=1/2$, $m_*=m_*^0=0.01$ m$^{-1}$, $\hat E=1 $. % ({\it Methods}~\ref{sec:6.2.1}).

We estimate these five parameters {\color{black}(using best fit to the functional form \eqref{eq:en-spec}, but either in frequency or in wavenumber space separately)} from (a) over 2000 independent timeseries from the Global Multi-Archive Current Meter Database (GMACMD), expanding the analysis of~\cite{le2021variability} to also include $s_{\rm NI}$ in addition to $s_\omega$ and $\hat E$ ({\it Methods}~\ref{sec:6.1.1}), and (b) 12 years worth of Argo float profiles, providing around $2.9$ million estimates of wavenumber slope $s_m$ and scale $m_*$\cite{pollmann2020global} ({\it Methods}~\ref{sec:6.1.2}). Fig.~\ref{fig:1} illustrates the geographic location of these estimates ({\bf A}) and
for select observed spectra
the fitting procedure to estimate  $s_{\omega}$ ({\bf B}), $s_{\rm NI}$ ({\bf C}), $\hat{E}$ ({\bf D}), and $s_m$ and $m_*$ ({\bf E}).

We use this combined information to build the largest possible dataset of 2D spectral estimates of the total internal wave energy ({\it Methods}~\ref{sec:6.2.2})
 -- hereafter referred to as the combined global dataset.
Figs.~\ref{fig:1}{\bf F}-{\bf J} show the histograms of the spectral parameters at those locations where all five are available after data interpolation, fitting and analysis, representing the observed global variability of the oceanic internal wavefield.

%%commented caveat
\begin{comment}
An important caveat is in order.
Both the FPs and the setup in the present manuscript rely on the assumption that most mixing is driven by internal waves. Their applicability is thus restricted to the open ocean interior \citep{polzin2014finescale}. The FP estimates of  $\epsilon$ 
%obtains $\epsilon$ estimates by comparing the observed internal wave field characteristics (e.g., shear, strain) with the GM spectrum. These estimates 
are known to be off by an order of magnitude or more at some specific locations where the processes leading to dissipation are not associated with  internal wave dynamics and perturb the wave-driven forward energy cascade \citep{watermanSuppressionInternalWave2014,klymakDirectBreakingInternal2008,cusack2020observed}. The areas where such conditions occur are located near the ocean boundaries where the boundary-layers dynamics (surface and bottom), the topographic interaction (at the bottom) provide paths for the energy to dissipate directly. Locations with strong submesoscale activities and the presence of fronts \citep{garabato2022kinetic} are also known to bias the FP estimations. This finestructure contamination issue and its impact on our results is examined in {\it Supplementary Information}. The field observations analyzed in this manuscript are restricted to the interior ocean excluding, as much as possible, the ocean's boundaries.
\end{comment}

\subsection*{Wave-wave interaction theory predicts mixing}\label{sec3}
A given test wave mode can exchange energy with a pair of other modes when their amplitudes multiply each other in the quadratic term of the primitive equations of the ocean~\cite{olbers2012ocean}. 
If nonlinearity is weak compared to linear dispersion, the energy transfers are dominated by resonant interactions, when the wavenumbers and frequencies of two waves sum respectively to the wavenumber and frequency of the third wave. \textcolor{black}{Such interactions in internal wavefields are observed at play in the ocean~\cite{mackinnon2013parametric}, in high-resolution models~\cite{mackinnon2005subtropical,skitka2023probing}, and in large water tanks~\cite{rodda2022experimental,lanchon2023internal}.} We treat this triadic resonant picture rigorously in the Hamiltonian formalism derived directly from the primitive equations~\cite{LT2} ({\it Methods}~\ref{sec:6.3.1}). We neglect interactions of internal waves with currents and mesoscale vortices, although these can modify a purely wave-driven forward energy cascade and the resulting energy dissipation~\cite{cusack2020observed,garabato2022kinetic,whalen2018large,chouksey2022gravity,wu2023reabsorption}, and e.g. the presence of vortices can impact the level and shape of the internal gravity wave continuum itself~\cite{danioux2015concentration,yang_oceanic_2023}.

\begin{figure}[H]%
\centering
\includegraphics[width=\textwidth]{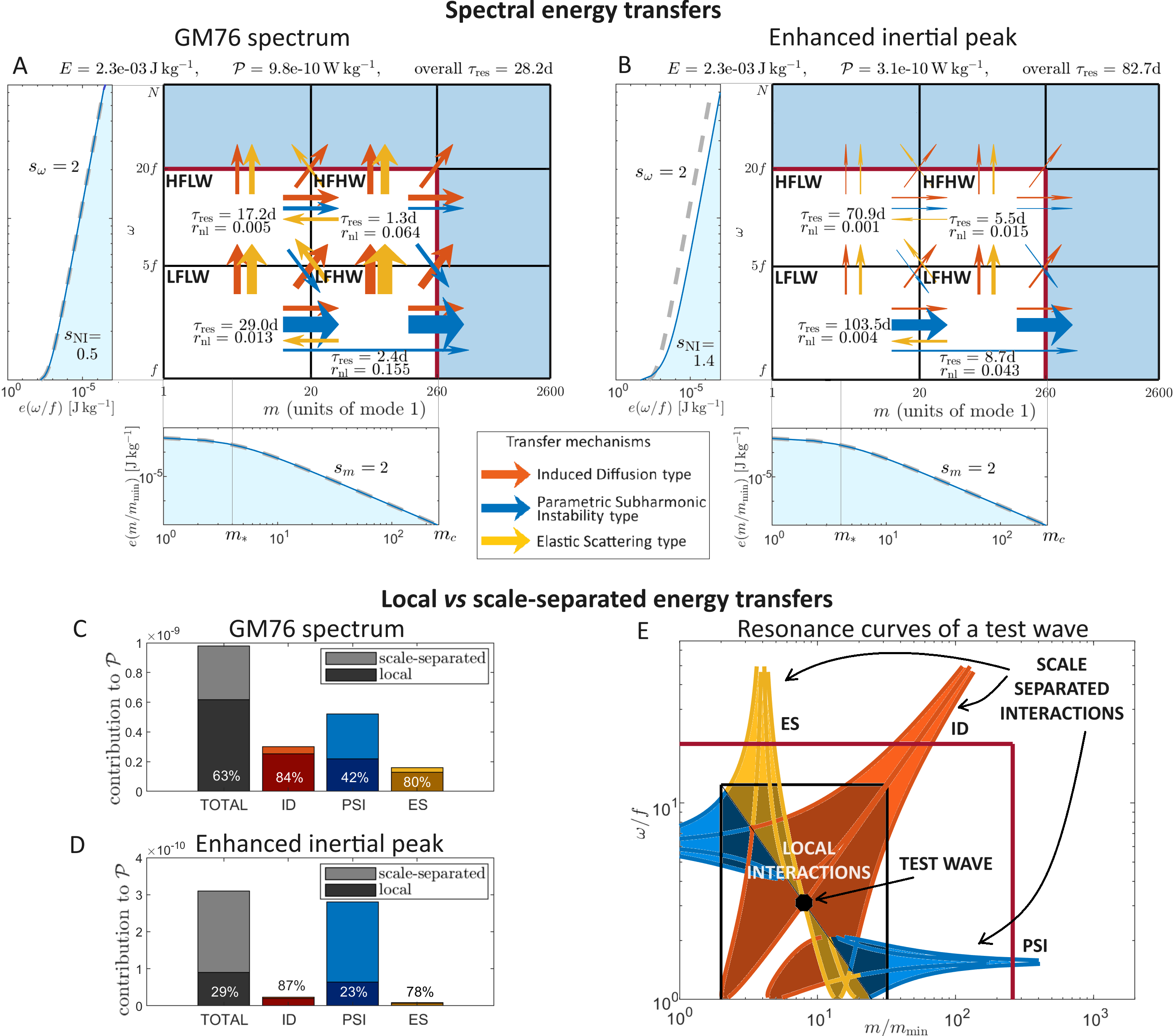}
\caption{{\bf Theoretical spectral energy transfers.} Prediction of spectral energy transfers for two exemplary spectra with same total energy $E$ and different near-inertial slope $s_{\rm NI}$. {\bf A}: Garrett-Munk model (GM76) with a shear-to-strain ratio $R_\omega=3$ (see~\eqref{eq:R_om} in {\it Methods}); {\bf B}: spectrum with larger near-inertial peak ($R_\omega = 7.3)$.
The frequency and wavenumber energy spectra are shown at the bottom and the left of panels {\bf A} and {\bf B}, with dashed grey lines representing the GM76 reference.
%The {\color{magenta}turbulent energy production} rate
$\mathcal P$ is the sum of the energy transfers across the red boundary separating the internal-wave band (white) from the turbulent dissipative region (light blue).  For each of the subregions we also report the nonlinear residence time $\tau_{\rm res}$ (ratio between the outgoing energy transfers and the energy contained in the subregion) and the nonlinearity parameter $r_{\rm nl}$ (ratio between the average wave period in the subregion and $\tau_{\rm res}$). We group the interactions into the three classes of {\color{black} Induced Diffusion (ID), Parametric Subharmonic Instability (PSI), and Elastic Scattering (ES) -- in spite of the original scale-separated definitions, here we keep track of the class also in the local regime (see panel {\bf E}).} {\color{black} The spectral subregions are labeled by the different combinations of high/low frequency (HF/LF) and high/low wavenumber (HW/LW)}. Panels {\bf C} and {\bf D} show (respectively for {\bf A} and {\bf B}) the contributions to $\mathcal P$ due to each of the mechanisms. Moreover, we quantify the proportion of such contributions that is due to spectrally local (vs scale-separated) interactions. {\color{black} Panel {\bf E} shows the energy transfers for a test wave, with the three branches and their respective local and scale-separated contributions (scale separation defined by a factor of 4 in $m-\omega$ space)}. 
}\label{fig:2}
\end{figure}

To quantify the energy transfers across the different scales, we partition the internal-wave band into nine subregions (Fig.~\ref{fig:2}), using a methodology for wave-wave energy transfers between arbitrary spectral subregions ~\cite{dematteis2023structure} ({\it Methods}~\ref{sec:6.3.1a}). This crucially differentiates our approach from other recent work~\cite{eden2019numerical}, allowing us to target quantitatively the locality and directionality of the spectral energy transfers. When energy is transferred to the dissipative regions, it is converted into turbulent energy ~\cite{garabato2022oceanBook}. Here, we quantify the turbulent energy production rate, $\mathcal P$, as the sum of the energy transfers leaving the internal-wave band toward the dissipative regions, as e.g. in~\cite{Muller86,eden2019numerical}. The result is a theoretical formula for $\mathcal P$, from which we are able to evaluate the vertical diffusivity, $K$, and the TKE dissipation rate, $\epsilon$, by use of a standard parameterization~\eqref{eq:4}. 
For $\epsilon$, the formula reads:

\begin{equation}\label{eq:2a}
	\epsilon = \epsilon_0^{\rm th}(s_{\rm NI},s_\omega,s_m) \frac{f}{f_0} \left(\frac{ E{m_*}^{s_m-1}}{  E_0{m_{*}^0}^{s_m-1}}\right)^2\,.
\end{equation}
An $N^2$ factor appears when using the normalized level $\hat E$. {\color{black} Moreover, in the case of $s_m=2$, corresponding to a white shear spectral density, we have that $E m_*$ scales like the shear-variance level $\mathcal E$, so that~\eqref{eq:2a} notably encompasses the scaling of~\eqref{eq:0} in terms of shear-variance level -- while a correction arises for $s_m\neq2$. {\color{black} In {\it Methods}~\ref{sec:6.3.2} we provide a detailed justification of this fact, }{\color{black} as well as the analytical relationship between our five spectral parameters and the normalized shear level of the GHP finescale parameterization~\cite{polzin2014finescale}.}  {\color{black}The factor encoding the dependence on the spectral slopes is fixed by  $\epsilon_0^{\rm th}= (1-R_f) \mathcal P_0^{\rm th}$, with $R_f=0.17$, using~\eqref{eq:4}, where $\mathcal P_0^{\rm th}$ is computed numerically \textcolor{black}{({\it Methods}} \ref{sec:6.3.2})}}.
The value associated with the original FP formula~\eqref{eq:0}, of $\mathcal P=8\times10^{-10}$ W kg$^{-1}$~\cite{polzin2014finescale}, is retrieved by our theoretical prediction {\color{black}when we use as input the canonical parameters of the GM76 spectrum, for which we obtain $\mathcal P=9.8\times10^{-10}$ W kg$^{-1}$. This is shown in Fig.~\ref{fig:2}A. In Fig.~\ref{fig:2}B, we then consider a different spectrum, with the same amount of energy redistributed with higher concentration in the inertial peak, and less energy in the high frequencies. We observe that this different spectral configuration depletes $\mathcal P$ considerably, by more than a factor of 3.  This happens because the frequencies close to $f$ are more linear than the high frequencies, resulting in an overall reduction of turbulent energy production. A difference in the energy pathways is also apparent in the comparison of Figs.~\ref{fig:2}A-B. We use these two cases as examples to show how different spectra imply different  pathways and magnitudes of energy transfer.}

 We further decouple the spectral energy transfers into the three primary classes of Induced Diffusion (ID), Parametric Subharmonic Instability (PSI), and Elastic Scattering (ES)~\cite{Muller86} -- see {\it Methods} \ref{sec:6.3.3} for more details. \textcolor{black}{For the GM76-type spectrum (Fig.~\ref{fig:2}A), the PSI class is responsible for roughly half of the total turbulent energy production, but this proportion varies for different spectra (Fig.~\ref{fig:2}B).}  
Originally, these three interaction classes were defined for the asymptotic regime of large scale separation ~\cite{McComas1977}. Previous works, including the original FP derivation~\cite{gregg1989scaling}, focus on such scale-separated interactions for estimating the energy transfers through the internal wave spectrum.  
{\color{black}Here, the resolution of the collision equation allows for an exhaustive evaluation of these three types of interaction classes to account for all interactions -- we consider an interaction local if all members of the triad are within a factor of 4 from each other (either in the $m$ or in the $\omega$ directions), and scale-separated otherwise}. For ID and ES we find that at least 80\% of the energy transfers are due to local interactions (Fig.~\ref{fig:2}C-D) {\color{black} -- with almost negligible scale-separated contributions, in agreement with the understanding that GM76 is an asymptotic ID and an ES zero-flux solution~\cite{dematteis2022origins,wu2023energy}}. Also for PSI the local contribution is substantial, {\color{black} and Figs.~\ref{fig:2}C-D show that the local {\it vs} scale-separated proportion depends strongly on the spectral distribution. This makes a difference, for instance, during the evolution of a spectrum initially forced at low frequency and low wavenumber. In such a case, we expect a transition between a regime characterized by interactions with large scale separation at the early stages of the evolution, and a significantly more spectrally-local regime at later stages when a stationary state is reached.}
In general, these findings contradict the notion that the scale-separated interactions represent the total energy transfer \citep{Muller86,gregg1989scaling} and highlight instead the role of the previously ignored interactions between waves of similar scales. This supports the results of \cite{onuki2018decay} to explain why there is such little observational evidence \citep{musgrave2022lifecycle} for the suggested catastrophic decline of internal-tide energy at the latitude where PSI is most efficient \citep{mackinnon2005subtropical}. {\color{black}In fact, this decline may not be as evident if the contribution by local interactions other than PSI is large and not particularly affected by the critical latitude itself.}

\subsection*{\color{black}Validation and comparison}\label{sec4}

{\color{black}First, we validate the theoretical formula~\eqref{eq:2a} with microstructure observations. Second, we apply~\eqref{eq:2a} to the combined global dataset of internal
wave spectra, compare it with strain-based FP estimates, and characterize the global distribution of mixing intensity and timescales.}

\subsubsection*{\textcolor{black}{Validation with microstructure measurements}}\label{sec4a}
\begin{figure*}[htb]%
\centering
\includegraphics[width=\linewidth]{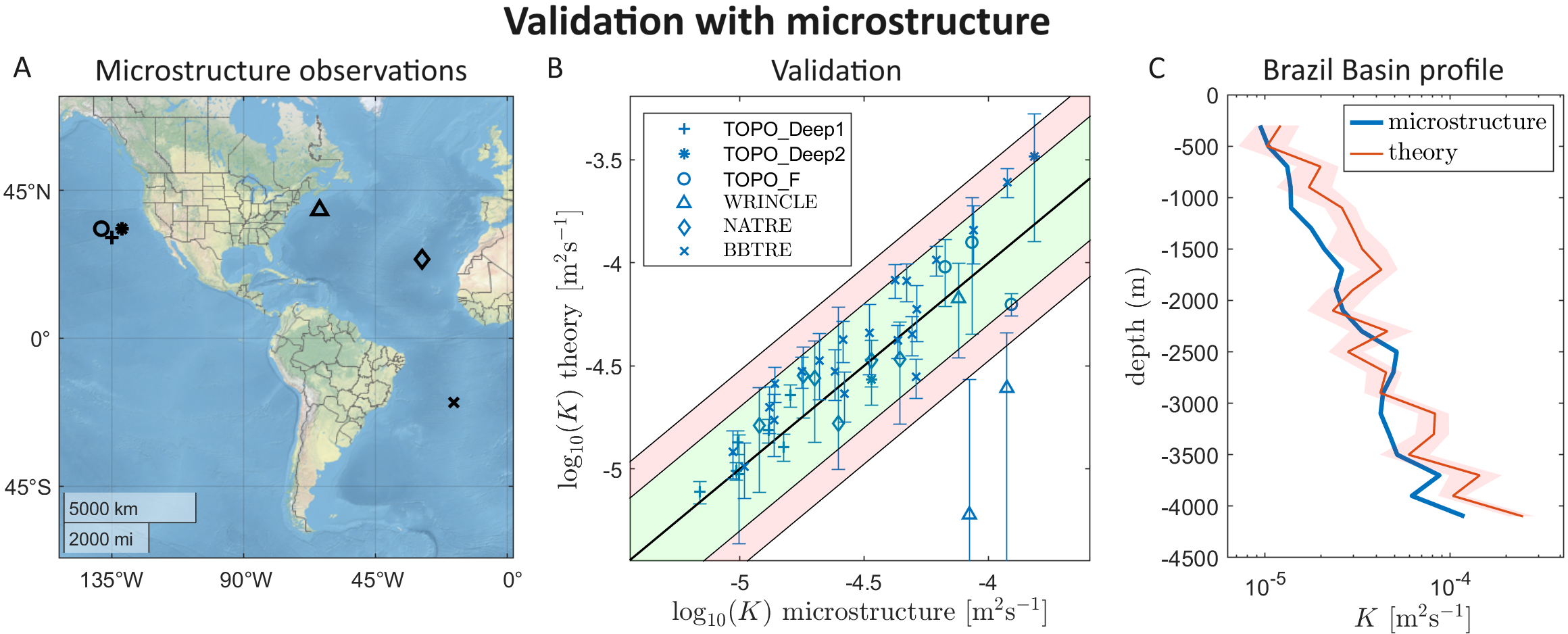}
\caption{%Evaluation and characteristics of our mixing estimates from first principles.
{\bf Validation with microstructure.} 
{\bf A}: Location of the High-Resolution Profiler (HRP) observations used for validation. {\bf B}: Validation of the theoretical calculation of vertical diffusivity $K$ with input from finestructure against the corresponding microstructure estimates, from simultaneous HRP measurements~\cite{polzin1995finescale}. %(section~\ref{sec4a}). % All but two points are in agreement by less than a factor of 2.
{\bf C}: Vertical profile of $K$ from the BBTRE observations conducted over rough topography~\cite{polzin1997spatial}. %The data points correspond to 20 overlying depth ranges of 200 meters each, resulting from the average of 30 full-column profiles.
}\label{fig:3}
\end{figure*}

We use time series sampled by the High-Resolution Profiler (HRP) in five distinct oceanographic regimes~\cite{polzin1995finescale} (Fig.~\ref{fig:3}{\bf A} and {\it Methods}~\ref{sec:6.1.3}): in the vicinity of a large seamount (Fieberling Guyot) in the eastern North Pacific Ocean; in a mid-ocean regime in the Atlantic; below a warm core ring of the Gulf Stream; {\color{black}and above the rough Mid-Atlantic Ridge in the eastern Brazil Basin.}  
These {\color{black}measurements} provide simultaneous independent access to micro- and fine-scales. We infer the 2D energy spectrum from the finescale shear and strain spectra {\color{black}({\it Methods}~\ref{sec:6.2.3}), and determine $K$ using our  theoretical formula for ${\mathcal P}$ and the empirical relation $K=R_f \mathcal P/N^2$ ({\it Methods}~\ref{sec:6.3.2}).} 

{\color{black}All except for two of 42 data points shown in Fig.~\ref{fig:3}{\bf B}, each the result of an average over several profiles, agree with the microstructure estimate of $K$~\cite{polzin1995finescale,polzin1997spatial} within a factor of 2 (slightly more in two cases).} %Another 3 points are in agreement with microstructure by a factor of 3.
The two  outliers were measured right below a strong warm core ring, where effects not considered here such as wave-mean flow interactions are expected to enhance mixing substantially~\cite{kunze1995energy}.  Considering the heterogeneity of the observations, the range of values spanning 1.5 orders of magnitude, and a nominal uncertainty of the microstructure estimates as high as $50\%$~\cite{polzin1995finescale}, the agreement is striking. 

{\color{black}For the BBTRE observations~\cite{polzin1997spatial}, each data point comes from the average of 30 full-column profiles, in one of 20 overlying depth ranges of 200 m. This allows us to show a vertical profile of turbulent diffusivity throughout the over 4000 m-deep water column in Fig.~\ref{fig:3}{\bf C}. The theoretical estimate with finestructure data input closely reproduces the microstructure vertical profile of (bottom-enhanced) mixing. Although we note an overall slight overestimate, this is comparable in size with the uncertainty on the theoretical estimate itself (shaded area in Fig.~\ref{fig:3}{\bf C}).}

\begin{figure*}[htb]%
\centering
\includegraphics[width=\linewidth]{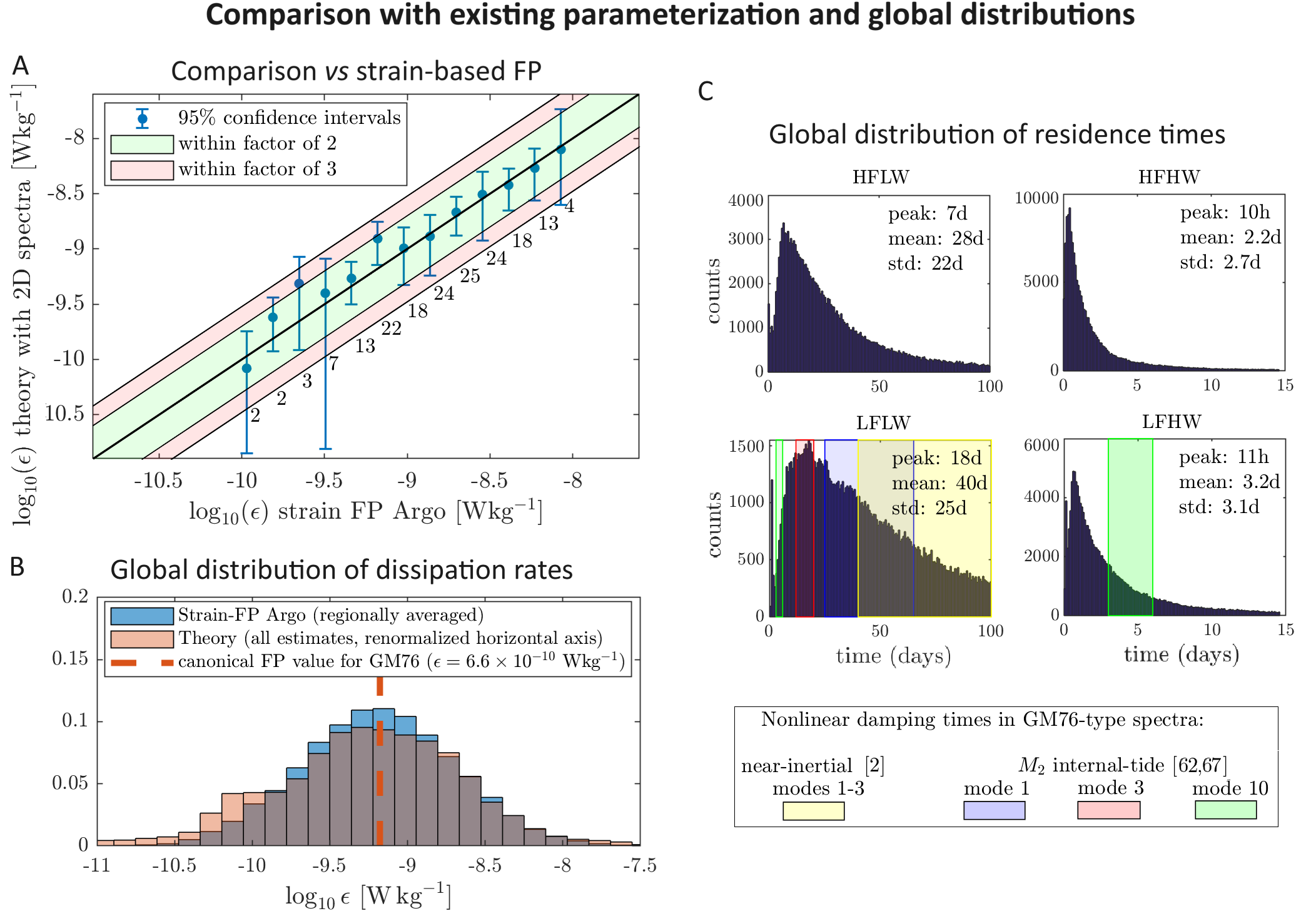}
\caption{%Evaluation and characteristics of our mixing estimates from first principles.
{\bf Comparison with existing parameterization and global distributions.} 
{\bf A}: Evaluations of regionally and temporally averaged $\epsilon$ in the global ocean, {\it vs} the strain finescale parameterization (FP) estimates of~\cite{pollmann2023}. %(section~\ref{sec4b}).
The data points where both evaluations of $\epsilon$ are available are bin-averaged along the horizontal axis (points per bin printed below errorbars).
{\bf B}: Distribution of all theoretical mixing estimates for the spectra in the global combined dataset shown in Fig.~\ref{fig:1}, compared with the updated global distribution of bin-averaged values of~\cite{pollmann2020global}. To compare on the same coarse-graining level, assuming statistical independence we renormalized the horizontal spread of the theoretical estimates, dividing by $\log_{10}\sqrt{\bar {n_i}}$, with $\bar {n_i}= 1023$ estimates/bin (average). %The canonical value of the FP formula for a GM76 spectrum, $\mathcal \epsilon = 6.6\times 10^{-10}$ Wkg$^{-1}$, is indicated by the red dashed line. 
{\bf C}: Histograms of coarse-grained nonlinear residence times at the different scales of global oceanic internal wavefields. At low frequency, the shaded regions are damping-time ranges of low-modes from previous literature (mode 10, at the boundary between low and high wavenumbers, shown in both panels). %-- notice that for the majority of the data points the two-sided $95\%$ confidence interval is not larger than a factor of 3.
}\label{fig:3b}
\end{figure*}

\subsubsection*{\textcolor{black}{Mixing patterns across the global ocean interior: comparison with an existing finescale parameterization.}}\label{sec4b}
We apply our generalized FP~(\ref{eq:2a}) to the combined global dataset of internal wave spectra and compare the results to the most up-to-date strain-based FP~(\ref{eq:0}) mixing estimates of~\cite{pollmann2023} from Argo float profiles. This reference uses the original strain-based FP incarnation of Eq.~\eqref{eq:0} \textcolor{black}{with a constant $\epsilon_0$}, which was validated against microstructure observations \citep{whalen2015estimating} and offers wider spatial coverage than the fully independent microstructure estimates used above. %in the evaluation of subsection~\ref{sec4a}.
%{\color{red} The authors of~\cite{ijichi2015frequency} raised concerns about potential bias in the strain-based FP for very large shear-to-strain ratios~\cite{ijichi2015frequency}. In {\it Methods}~\ref{sec:6.2.2}, we show that the average shear-to-strain ratios in our dataset are always smaller than 10. Thus, our results should not be affected by their correction~\cite{ijichi2017eikonal}.}
We sort and average our estimates in the 1.5$^{\circ}\times$1.5$^{\circ}$ horizontal bins and the three depth ranges ($250$m$-500$m, $500$m$-1000$m, and $1000$m$-2000$m) of the Argo-based reference.

\begin{figure}[ht]%
\centering
\includegraphics[width=\columnwidth]{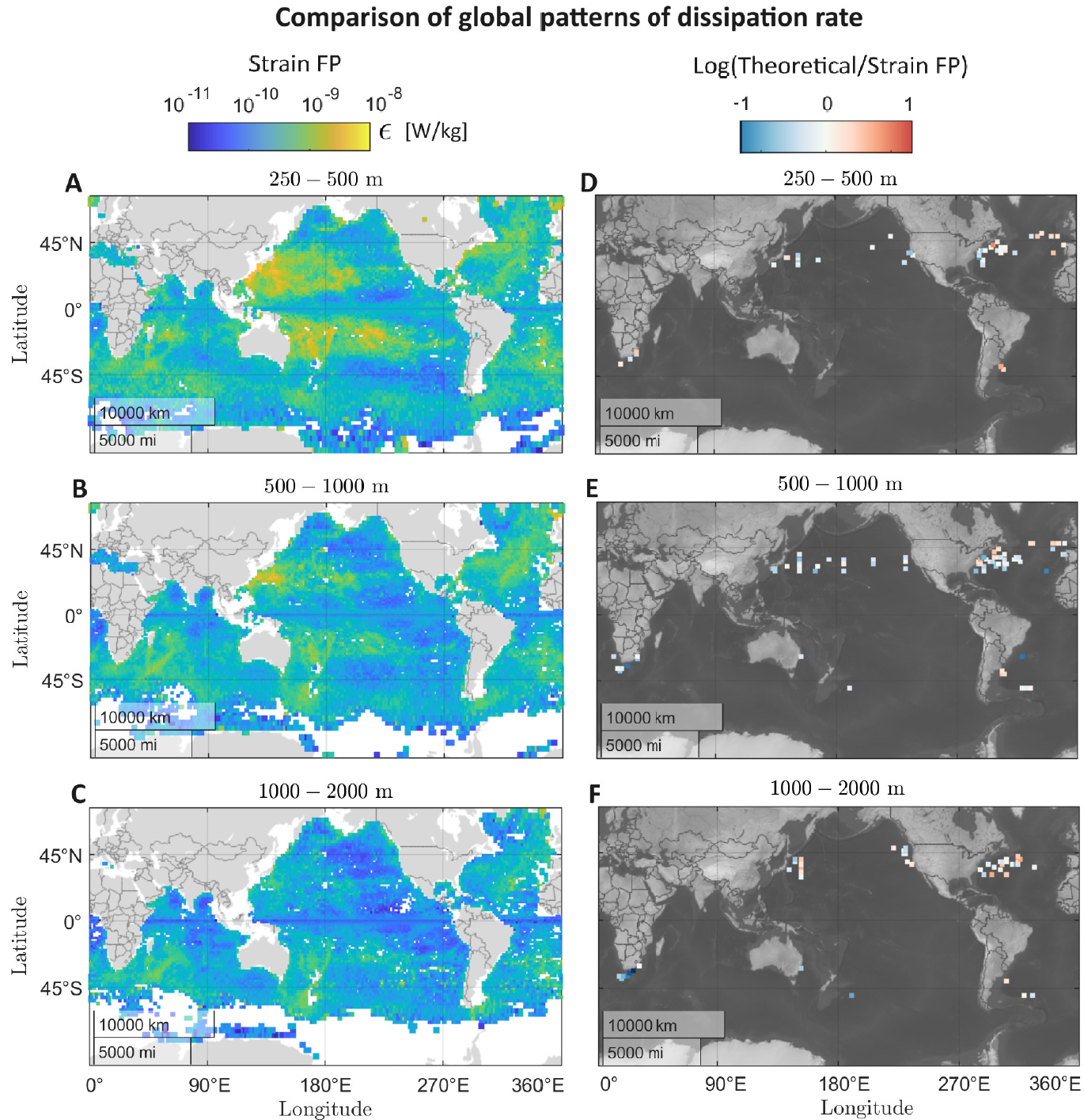}
\caption{{\bf Comparison of geographic distributions of turbulent kinetic energy dissipation rate $\epsilon$.} Visual comparison between the theoretical estimates of dissipation rate using our generalized finescale parameterization (FP) Eq.~(2) with input from the global combined dataset, and the strain FP estimates with input from Argo data. Panels {\bf A}-{\bf C} show the Argo reference in three different depth levels, while panels {\bf D}-{\bf F} show the ratio between the theoretical estimate and the Argo reference, respectively for the three depth levels. Shown are the average values obtained over the regional bins of 1.5$^\circ\times$1.5$^\circ$ horizontally, and in the three given depth layers vertically. These data points are used to build the plot in Fig.~\ref{fig:3b}{\bf A}. In the right panels, white color indicates agreement, while red and blue color indicate respectively a higher and lower theoretical value with respect to the strain FP value.}\label{fig:9M}
\end{figure}

We find a total of 175 bins for which both our new (average 1023/bin) and reference (average 740/bin) estimates are available.  We find highly significant agreement over two orders of magnitude (Fig.~\ref{fig:3b}{\bf A}). %, with almost all points including agreement (main diagonal) within their 95\% confidence interval.
This {\color{black}strengthens} the success of our generalized FP (\ref{eq:2a}) {\color{black}validated} in the previous subsection. It also implies that for present-day conditions, the original FP incarnation with an empirically set constant $\epsilon_0$ (\ref{eq:0}) provides comparable estimates to our generalized FP (\ref{eq:2a}). However, for climate conditions that differ substantially from today's, only a parameterization based on the underlying physics alone like our generalized FP (\ref{eq:2a}) can ensure reliable mixing estimates.

The distribution of all our calculated values of $\epsilon$ collapses onto the distribution of the strain-based FP estimates of~\cite{pollmann2023} (Fig.~\ref{fig:3b}{\bf B}),  once the spread of the theoretical estimates is suitably renormalized to account for regional coarse-graining. We now define the {\color{black}coarse-grained} residence time, % due to the nonlinear wave-wave interactions,
$\tau_{\rm res}$, as the ratio between the total energy in a given subregion and the sum of its outgoing energy transfers. %This is an integral analog of the time scales proposed, for instance, in.
%A point-wise version~\cite{holloway1980oceanic,mccomas1981time} is studied in {\it Supplementary Information}~\ref{sec:6.3.4}.
Fig.~\ref{fig:3b}{\bf C} depicts the broad distributions of our predicted residence times, with mean values from a few days (high wavenumbers) to about $30$ days (low wavenumbers).
This multiscale character is reminiscent of the dual treatment of the internal tide in the OGCMs: the low-wavenumber far field, with large residence times allowing for hundreds of kilometers of propagation from the source, and  the high-wavenumber near field, dissipated much faster near the generation site~\cite{garabato2022oceanBook}. In the low-frequency panels, we show the range of wave-wave damping timescales for selected modes in GM76-type spectra: from the very weakly nonlinear low-mode near-inertial waves~\cite{Muller86} (up to 100 days), to the $M_2$ baroclinic modes spanning from 50 days (mode 1) down to a few days (mode 10)~\cite{onuki2018decay,olbers2020psi}.  We find general consistency of our predicted timescales with prior knowledge, whilst also providing a notion of typicality of nonlinear timescales in observed oceanic conditions.

Fig.~\ref{fig:9M} shows a geographical comparison between
our bin-averaged theoretical estimates of TKE dissipation rate (with input from the global combined dataset) and the bin-averaged estimates from strain FP of~\cite{pollmann2023}. 
%Figure~\ref{fig:7M} shows a geographical comparison between our regionally and temporally averaged predicted dissipation rates and the strain-based predictions of~\cite{pollmann2023}.
Agreement is generally strong, although a few locations show some disagreement -- a region-dependent study of the reasons of disagreement, which will necessarily include a combined analysis of the energy sources at play, is beyond the scope of the present manuscript and will be addresses in future work.

A study of nonlinearity level reported in {\it Supplementary Information} shows that, contrary to earlier arguments~\cite[e.g.][]{holloway1980oceanic}, oceanic internal waves are typically weakly nonlinear at most scales. The scales that can involve lower residence times and higher nonlinearity levels concern the low-frequency, high-wavenumber PSI decay. Strong nonlinearity, indeed, is expected as the wavebreaking scales are approached.
%This, however, is required to connect to the inherently nonlinear wave breaking and turbulence regimes and hence to be expected.

\section*{Discussion}
{\color{black}For the sake of transparency, we underscore the constraints inherent in our approach: we (i) overlook processes other than wave-wave interactions within a spatially homogeneous, vertically symmetric and horizontally isotropic wavefield; (ii) employ the exactly resonant formalism of weak wave turbulence theory {\color{black}implying, among other things, that we neglect the advective ``sweeping'' of small-scale waves by large-scale waves via the Doppler effect~\cite{lvov2024generalized,polzin2022one};} (iii) operate under the hydrostatic balance assumption; (iv) assume separable spectra -- indeed an oversimplification. Moreover, (v) the weakly nonlinear assumption is arguably broken near the high-wavenumber wave-breaking threshold $m_c$, but the results are rather insensitive of the choice of $m_c$; and (vi) an assumption of regionality and stationarity underlies our cross-matching of internal-wave spectra databases.} {\color{black}The above points define the limitations of our approach. In particular, it cannot describe the dynamical transitions to instabilities or other non-weak phenomena that are a necessary bridge from weakly nonlinear theory to 3D turbulence, such as caustics in ray theory (related to boundary layer theory), bound waves, and transient non-resonant interactions that represent a transition to wave instability or stratified turbulence~\cite{broutman1986internal,broutman1986interaction,d2000wave,rodda2023internal}. This is a topic of intense investigation from theoretical, numerical, experimental, and observational perspectives.}

{\color{black}Our results are, at least qualitatively, in line with results from high-resolution numerical simulations, in particular regarding a high dependence on spectral shape and the large contribution from local interactions~\cite{furue2003energy,pan2020numerical,skitka2023probing}. On the one hand, we refrain from comparing directly with previous numerical simulations restricted to a 2D vertical plane~\cite[e.g.][]{hibiya1996direct,bel2022resonant}, expecting the lower dimensionality to change the properties of the system by reducing linear dispersion and thereby increasing the effective nonlinearity. On the other hand, future work will need to address the comparison between different methodologies of quantification of spectral energy transfers, including triad estimates and coarse-graining methods~\cite{dematteis2023structure,skitka2023probing,barkan2021oceanic,eyink2009localness}.}

{\color{black}Within the framework of assumptions listed above, we have demonstrated that our theoretical/numerical evaluations of turbulent mixing come straightforwardly from the dynamical equations of the ocean, and are free of arbitrary parameters that could be used heuristically to fine-tune the estimates. In this sense, our quantification is from first principles in essence. Considering the apparently strong operational assumptions, it is all the more striking that our  first-principle quantification of turbulent mixing (which acts as a theoretical/ conceptual generalization of the FP formula), with input from observational spectra, is in such strong agreement with widespread observations in the global ocean interior -- both with direct microstructure measurements at some select locations~\cite{polzin1995finescale}, and with the most up-to-date corroborated indirect estimates in the global ocean~\cite{pollmann2023}, via strain-based FP applied to Argo data.} This demonstrates the close causal link between wave-wave interactions,  particularly the spectrally-local ones that were previously ignored, and turbulent mixing. This assessment paves the way for many further analyses, such as a robust study of the spatio-temporal variability of internal wave energy transfers, wave-induced mixing, and the link to environmental conditions. Such information is essential for the design of experiments or research cruises as well as the development of energy-constrained mixing parameterizations, which are indispensable for the reliable modeling of future changes in ocean dynamics and their global-scale impact \cite{olbers2019idemix,melet2022role,zhang2021decreased}.

\bigskip

\section*{Methods}\label{sec11}
\appendix
%Topical subheadings are allowed. Authors must ensure that their Methods section includes adequate experimental and characterization data necessary for others in the field to reproduce their work. Authors are encouraged to include RIIDs where appropriate. 

%We can use up to 10 additional figures that can be referenced in the Methods, and the Methods can be up to 3000 words total (longer than main text that roughly should be limited to 2500 words and 4 display items).

\section{Field data}\label{sec:6.1}
\subsection{Global Multi-Archive Current Meter Database}\label{sec:6.1.1}

Ocean velocities timeseries were selected from the Global Multi-Archive Current Meter Database (GMACMD). Following \cite{le2021variability}'s methodology, 2260 current meters on 1362 moorings were selected (Fig. \ref{fig:1}). The instruments’ depths range from 100 m below the surface to 200 m above the seafloor with a maximum of 6431 m depth. Finally, data within 5° of the equator were discarded to avoid complications with the longer inertial periods at very low latitudes.

The time evolution of kinetic energy frequency spectra ($\phi_{KE}(\omega)$) for each of these time series is evaluated using a multitaper method with a sliding 30-day window. %The observed spectral kinetic energy level is normalized by the default GM76 buoyancy frequency (3 cph) in order to compare the data to the GM76 spectrum.
After removing the spectral peaks associated with the main energy sources \textcolor{black}{(removing frequency bands around the near-inertial \textcolor{black}{and} tidal peaks \textcolor{black}{as well as} their harmonics)}, the slope ($s_\omega$) of the internal wave continuum is estimated with a linear fit. A similar method is used to estimate the slope ($s_{\rm NI}$) of the near-inertial peak using $\phi_{KE}(\frac{\omega-f}{f})$.

\subsection{Argo floats}\label{sec:6.1.2}
Argo floats autonomously profile the ocean's upper 2000\,m, collecting temperature, salinity and pressure information at a vertical resolution of a few meters roughly every 10\,days \cite[e.g.][]{wong2020argo}. Building on previous implementations of the finestructure parameterization \cite{kunze2006global, whalen2012spatial, pollmann2017evaluating}, \cite{pollmann2020global} derived vertical wavenumber spectra of strain $\xi_z = (N^2-N^2_{fit})/\overline{N^2}$, where $N^2$ is the buoyancy frequency, $N^2_{fit}$ a quadratic fit and $\overline{N^2}$ the vertical mean, from these profiles, translated them into energy spectra by exploiting the polarization relations \cite{olbers2012ocean}, and fitted the GM76 model to obtain a global data base of internal wave energy level, vertical wavenumber slope $s_m$ and wavenumber scale $m_*$, which we here use and are available at \url{https://doi.org/10.5281/zenodo.6966416} \citep{pollmann2022}. The integrated strain spectrum provides the strain variance, a key component of the FP expression to estimate TKE dissipation rates. We here use an update of the TKE dissipation rate estimates from Argo float profiles  of \cite{pollmann2017evaluating}, available at \url{ https://doi.org/10.17882/95327} \citep{pollmann2023}. %Details 

\subsection{High-Resolution Profiler}\label{sec:6.1.3}

The free-falling, internally recording HRP~\cite{schmit1988development} samples the water column giving access to spectra in vertical wavenumber space. It features an acoustic velocimeter and a Conductivity-Temperature-Depth (CTD) probe, from which we obtain the finestructure spectra of shear and strain variance, respectively. Independently, the HRP also has airfoil probes to sense the centimeter-scale velocity field. This makes it possible to reconstruct the microstructure shear variance, yielding a direct estimate of the TKE dissipation rate $\epsilon$ and, by use of the relations ~\eqref{eq:4}, of the diffusivity $K$.

The  {\color{black}five observational campaigns} that we use are: TOPO\_Deep, in the eastern North Pacific Ocean, consisting of eight profiles as deep as 3000 m; TOPO\_F, consisting of 15 profiles near TOPO\_Deep but with enhanced non-GM characteristics; the 40 WRINCLE  profiles from the center of a warm core ring of the Gulf Stream; 10 profiles from the NATRE experiment, in the North Atlantic mid-ocean regime; {\color{black}30 profiles from the BBTRE experiment~\cite{polzin1997spatial}, conducted above the rough topography of the Mid-Atlantic Ridge in the eastern Brazil Basin}. All profiles used in this analysis start at least 200 m above the bottom.
Each of the 42 independent data points represented in Fig.~\ref{fig:3}{\bf B} roughly corresponds to a one-day average of multiple vertical profiles that extend in depth from $1000$ m in the WRINCLE experiment to over $4000$ m in the BBTRE experiment. When they belong to the same experiment, the data points come from well-separated conditions in space or time, ensuring complete statistical independence. Detailed information on the five {\color{black}observational campaigns} is found in~\cite{polzin1995finescale,polzin1997spatial}.

\section{Data processing: from observations to 2D spectra}

\subsection{Adaptive formula for the 2D spectral energy}\label{sec:6.2.1}

Due to the large aspect ratio of the horizontal-to-vertical scales of oceanic internal waves, a minimal statistical description of the wavefield -- assuming no preferential directionality in the horizontal plane -- must be two-dimensional (2D). We represent the spectra in frequency ($\omega$) and magnitude of vertical wavenumber ($m$) -- the natural dual variables for fixed-point time series and vertical stratification profiles, respectively.
{The internal-wave band is bounded in frequency space by the local Coriolis frequency ($f$) and the buoyancy frequency ($N$), and in the vertical wavenumber space by  the mode-1 wavenumber $m_0=\pi/H$ where $H$ is the full water column, and the wavenumber of the largest available mode that is not subject to shear instability~\cite{garabato2022oceanBook}: 

\begin{equation}
m_c=2\pi/\ell_c
\label{eq:mc}
\end{equation}
 where $\ell_c = \sqrt{2{\rm Ri}_c KE}/N={\rm O}(10 $m$)$ is related to the critical Richardson number of Kelvin-Helmholtz instability ${\rm Ri}_c=1/4$, and $KE$ is the kinetic energy density per unit of mass.} %In this rectangular region in spectral space, the dominant share of the ocean's total energy is in the internal wave field.

We model the total energy spectral density with the following five-parameter formula (separable in frequency and wavenumber),

\begin{equation}\label{eq:en-spec}
	e(m,\omega) = B \frac{\omega^{2s_{\rm NI}-s_\omega}}{(\omega^2-f^2)^{s_{\rm NI}}}\frac{1}{m^{s_{ m}} + m_*^{s_{ m}}}\,,
\end{equation}
where $B$ is a normalization constant constrained by the total energy $E$.
In order for $E$ to always be finite, we assume a regularization plateau for frequency smaller than $1.025f$, such that $e(m,\omega<1.025f) = e(m,1.025f)$. {\color{black}Our justification of the plateau width is twofold: first, it is of the order of the frequency space resolution implied by our observational sliding window of 30 days ({\it Methods}~\ref{sec:6.1.1}), at mid latitudes; second, we observe that most energy close to $f$ (in the plateau region) is ``frozen'' in an effectively noninteracting linear state that does not affect the nonlinear energy transfers quantified in this manuscript ({\it cf. Supplementary Information} Fig. S2, showing residence times larger than 100 days in the inertial limit).  }
The choices $s_{\rm NI}=1/2$, $s_\omega=2$, $s_m=2$, $E=2.3\times 10^{-3}$ J$\,$kg$^{-1}$ (instead of $E=3\times 10^{-3}$ J$\,$kg$^{-1}$ because of the regularization plateau at $f$), $m_* = 4\pi/b$ (with $b=1300$ m) evaluated at $N_0=3$ cph, $f_0=2\sin({\rm lat})$ cpd, at latitude lat$\,=32.5^\circ$, reduce formula~\eqref{eq:en-spec} to the reference GM76 spectrum. The following sections provide details on the estimate of the five spectral parameters in Eq.~\eqref{eq:en-spec} from observational data.

\subsection{Combined global dataset}\label{sec:6.2.2}
\begin{figure}[ht]%
\centering
\includegraphics[width=\columnwidth]{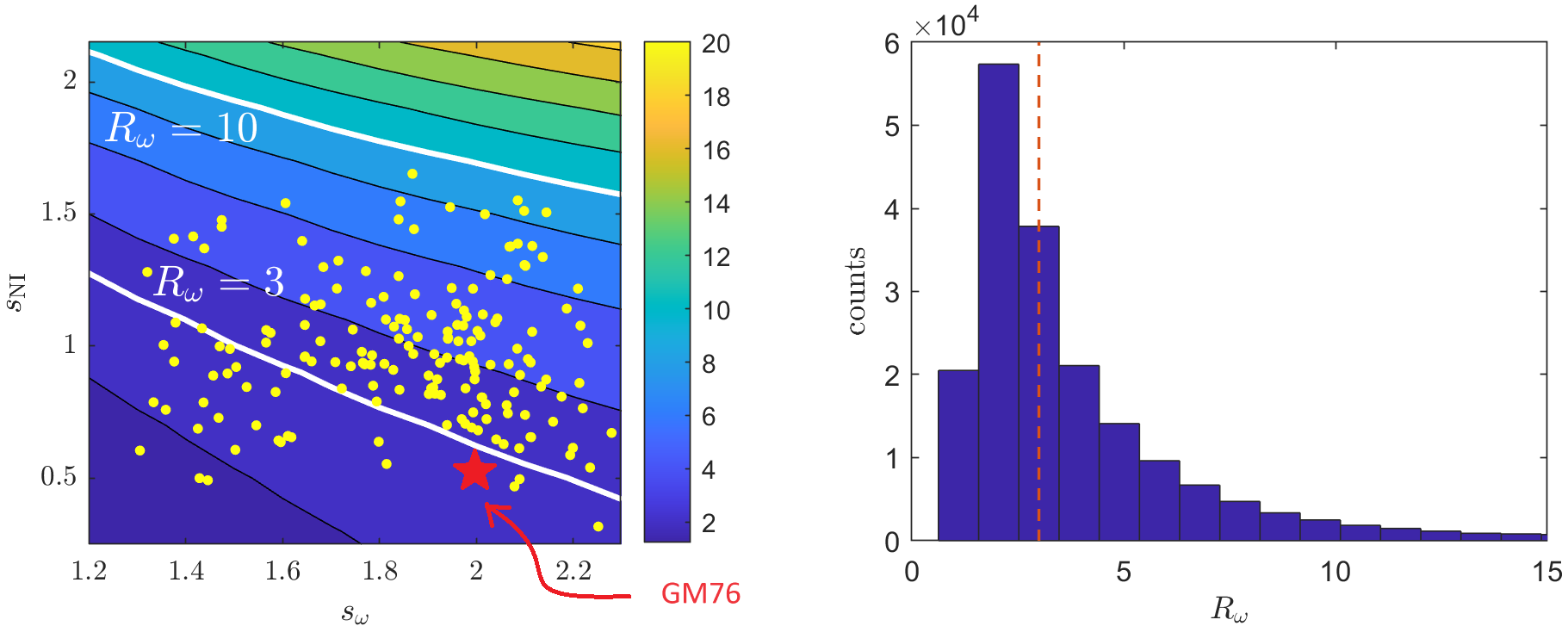}
\caption{{\bf Shear-to-strain ratio.} Left: Color map representing the shear-to strain ratio $R_\omega$ as a function of the spectral parameters $s_\omega$ (high-frequency slope) and $s_{\rm NI}$ (near-inertial
slope), according to formula (5). {\color{black}The yellow dots represent the spread of bin-averaged slopes from the 175 regions in the global combined dataset}. Right: Global distribution of $R_\omega$. The red
dashed line indicates the Garrett-Munk spectrum (GM76) value of $R_\omega$ = 3.}\label{fig:5M}
\end{figure}

In our global analysis of 2D internal-wave spectra, the three parameters $s_{\rm NI}$, $s_\omega$ and $E$ are estimated for each moored frequency spectrum of kinetic energy computed as described in section~\ref{sec:6.1.1}. By the polarization relations of internal waves~\cite{regional}, the near-inertial spectrum of total energy (and therefore $s_{\rm NI}$) is dominated by kinetic energy, \textcolor{black}{since the near-inertial waves are mostly horizontal-velocity oscillations}. On the other hand, the high-frequency slope $s_\omega$ is asymptotically the same for the kinetic and potential energy. The total energy $E$ relates to the kinetic energy $KE$ {\color{black}(neglecting vertical kinetic energy)} via

\begin{equation}
    E = KE \frac{R_\omega + 1}{R_\omega}\,,
\end{equation}
where the shear-to-strain ratio $R_\omega$ {\color{black}(which equals the ratio of horizontal kinetic energy to available potential energy)}, assuming separability, is computed as

\begin{equation}\label{eq:R_om}
    R_\omega(s_{\rm NI},s_\omega) = \frac{1}{N^2} \frac{\int_f^N \frac{N^2-\omega^2}{N^2-f^2}\frac{\omega^2+f^2}{\omega^2}\frac{\omega^{2s_{\rm NI}-s_\omega}}{(\omega^2-f^2)^{s_{\rm NI}}}  d\omega}{\int_f^N \frac{\omega^2-f^2}{(N^2-f^2)\omega^2}\frac{\omega^{2s_{\rm NI}-s_\omega}}{(\omega^2-f^2)^{s_{\rm NI}}} d\omega}\,,
\end{equation}
as represented in Fig.~\ref{fig:5M}~(left) {\color{black}(see e.g.~\cite{pollmann2020global})}. {\color{black}The yellow data points show the spread of the bin-averaged spectral slopes, with values of $R_\omega$ always smaller than 10. This fact is important as it justifies a-posteriori the applicability of the strain-based FP in these regions, in light of the correction proposed by~\cite{ijichi2015frequency,ijichi2017eikonal}, which would be relevant only for larger values of $R_\omega$.} The global distribution that we obtain for $R_\omega$ is shown in Fig.~\ref{fig:5M}~(right).

We complete the set of spectral parameters by assigning the average vertical wavenumber slope $s_m$ and scale $m_*$ computed from the Argo dataset (cf. section~\ref{sec:6.1.2}) for the discretization bin that contains the given moored spectrum.
Two assumptions of statistical regionality and stationarity underlie our analysis. The results of~\cite{regional} indicate that the regional characterization of spectral power laws after averaging over the mesoscale-eddy space-time scales is fairly independent of the seasonal variability of $E$.
Closer spatial correlation and seasonal variability will be analyzed in a follow-up study.

\subsection{High-resolution profiler observations}\label{sec:6.2.3}

\begin{figure*}[ht]%
\centering
\includegraphics[width=\textwidth]{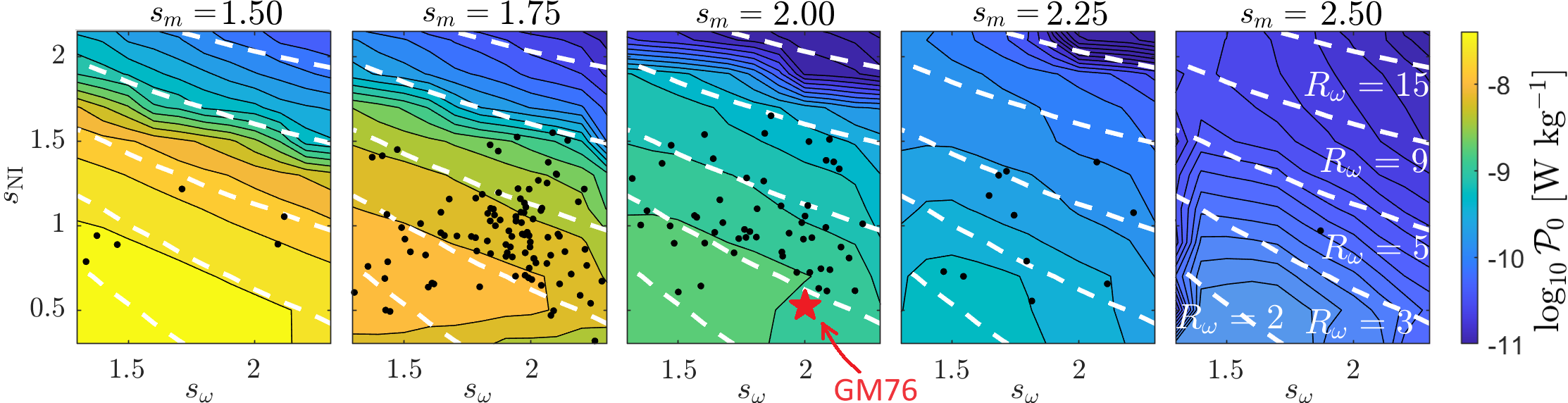}
\caption{{\bf Numerical evaluation of theoretical parameterization.} Color map of the theoretical reference turbulent prodution rate $\mathcal P_0^{\rm th}(s_{\rm NI},s_\omega,s_m)$, function of the three spectral slopes (near inertial $s_{\rm NI}$, high-frequency $s_\omega$, and high wavenumber $s_m$), computed numerically on a grid of $5\times5\times5$ points (cf. {\it Methods}~C.3). {\color{black}The dashed white lines are level sets of shear-to-strain $R_\omega$. When only the shear-to-strain ratio is available, $s_\omega$ and $s_{\rm NI}$ are not known. In that case, we exploit the fact that the level sets of $P_0^{\rm th}$ and $R_\omega$ are nearly parallel. We assign to our estimate the value of $P_0^{\rm th}$ intercepted on the fixed level set of $R_\omega$ at $s_\omega=2$, and evaluate the uncertainty error from the observed variability of $P_0^{\rm th}$ along the level set of $R_\omega$. We find that this uncertainty is always smaller than $10\%$ for the observed values of $R_\omega$}.}\label{fig:6M}
\end{figure*}

From the HRP profiles we obtain the finestructure spectra of shear and strain variance. We obtain directly the vertical wavenumber slope $s_m$ and scale $m_*$, and the total energy $E$ by use of the polarization relations and by direct calculation of $R_\omega$ as the shear-to-strain ratio. Now, we need to infer information on the frequency slopes $s_{\rm NI}$ and $s_\omega$. To this end, we exploit the fact that $R_\omega$ depends solely on the frequency spectrum~\eqref{eq:R_om}, since the wavenumber spectral properties cancel out in the ratio. Therefore, we have that $R_\omega=R_\omega(s_{\rm NI},s_\omega)$, as shown in Fig.~\ref{fig:5M}. A given value of $R_\omega$ therefore corresponds to a level set establishing a unique curve in the $s_{\rm NI}-s_\omega$ plane. For an evaluation of {\color{black}turbulent energy production} rate ($\mathcal P$), we study the variation of the factor $\left. \mathcal{P}_0^{\rm th}(s_{\rm NI},s_\omega,s_m)\right\vert_{s_m}$ in formula~\eqref{eq:3}, at fixed $s_m$, along the given $R_\omega$ level set in the $s_{\rm NI}-s_\omega$ plane. The resulting variation turns out to be small as the level sets of $\left. \mathcal{P}_0^{\rm th}(s_{\rm NI},s_\omega,s_m)\right\vert_{s_m}$ are mostly parallel to those of $R_\omega(s_{\rm NI},s_\omega)$, as can be appreciated in Fig.~\ref{fig:6M}. The uncertainty associated with this variation is smaller than $10\%$ in most regions of the  $s_{\rm NI}-s_\omega$ space. For operational purposes, we are thus free to fix a value of $s_\omega$ (for simplicity, a choice of $s_\omega=2$ is made here), %for the results in section~\ref{sec4a}), 
and determine the value of $s_{\rm NI}$ constrained by the computed value of $R_\omega$. The ensuing mixing evaluation is independent of the choice up to a small uncertainty.

\section{Theoretical methods}

\subsection{Hamiltonian formalism of internal wave-wave interactions}\label{sec:6.3.1}

 We start from the simple Boussinesq approximation of the primitive equations of the ocean ~\cite{olbers2012ocean,vallis2017atmospheric}, and use the Hamiltonian formalism of~\cite{LT2}. Assuming hydrostatic balance, the dispersion relation of internal waves is given by
 
\begin{equation}\label{eq:disp}
\omega^2 = f^2 + \frac{g^2}{\rho^2N^2}\frac{k^2}{m^2}\,,
\end{equation}
where $k$ and $m$ are the magnitudes of the horizontal and vertical components of wavenumber, $\rho$ is the mass density per unit of volume, and $g$ is the acceleration of gravity. 
%\begin{comment}
{\color{black}
We now use isopycnal coordinates ($x,y,\rho$) and assume spatial homogeneity and constant potential vorticity on each isopycnal layer. {\color{black}For simplicity of notation, here and in the rest of section C, $m$ denotes the vertical wavenumber in isopycnal coordinates.} This reduces the equations to two field variables, the velocity potential $\phi(x,y,\rho) $ (such that the horizontal velocity is given by ${\bf u}=\nabla \phi + \nabla^\perp \psi$, {\color{black}where $\nabla^\perp = (-\partial/\partial_y,\partial/\partial_x)$}), and the differential layer thickness $\Pi(x,y,\rho) = \rho \partial z /\partial \rho$. The divergence-free field $\psi$ is dynamically constrained to $\Pi$ by the constant potential vorticity assumption. The two $\phi(x,y,\rho) $ and $\Pi(x,y,\rho)$ are Hamiltonian canonically conjugated variables with respect to the Hamiltonian  

\begin{equation}
%\begin{aligned}
\mathcal H = \frac12 \int dxdyd\rho \left( \Pi \vert\nabla \phi + \nabla^{\perp}\psi\vert^2
\quad - g \left( \int_0^\rho \int_0^{\rho_1} d\rho d\rho_1 \frac{\Pi(\rho_2) - \bar\Pi(\rho_2)}{\rho_2} \right)^2 \right)\,,
%\end{aligned}
\end{equation}
where $\bar\Pi(\rho) = \rho \partial \bar z(\rho)\partial \rho = -g/N^2$ is the given stratification in hydrostatic balance (here assumed constant by the vertical homogeneity assumption). This means that the equations of motion take the simple form:

\begin{equation}
\frac{\partial \Pi}{\partial t} = \frac{\delta \mathcal H}{\delta \phi}\,,\qquad \frac{\partial \phi}{\partial t} = - \frac{\delta \mathcal H}{\delta \Pi}\,.
\end{equation}
 The Hamiltonian $\mathcal H$ has a cubic nonlinearity in addition to harmonic terms.
The harmonic part is diagonalized by the canonical normal variables $a({\bf p}) = N/k \sqrt{\omega({\bf p})/(2g)} \Pi({\bf p}) - i k/N \sqrt{g/(2\omega({\bf p}))} \phi({\bf p}) $, where we have now switched to Fourier space and ${\bf p}$ denotes the 3D wavevector.

We now use the relationship $\langle a({\bf p}) a^*({\bf p}') \rangle  = n(\bf p)\delta({\bf p} - {\bf p'})$ (the symbol $*$ denotes complex conjugation and $\langle\cdot\rangle$ denotes statistical average over a suitable statistical ensemble) defining the 3D wave-action spectrum $n(\bf p)$ in spatially homogeneous conditions. The 3D energy density is given by $e({\bf p})=\omega({\bf p})n({\bf p})$, with respect to which the 2D energy density is given by $e(m,\omega) = 4\pi (g/(\rho_0 N))^2 m^2\omega e(\bf p) $,} {\color{black}with $\rho_0=10^3$ kg/m$^3$. }
%\end{comment}

Following the standard approach of wave turbulence~\cite{hasselmann1966feynman,VLvov1992,NazBook}, 
%assuming {constant potential vorticity on layers of constant density, and using isopycnal coordinates}, 
under the assumption of weak nonlinearity, in a random near-Gaussian wavefield with spatially homogeneous, {\color{black}horizontally isotropic and vertically symmetric} statistics, we derive the Wave Kinetic Equation  for the spectral energy density~\cite{olbers1976nonlinear,LT2}:

\begin{equation}\label{eq:1M}
	\dot e(m,\omega) = I[e,e](m,\omega)\,,
\end{equation}
where $I[e,e]$ is the wave-wave collisional operator, a quadratic integral operator that accounts for the contributions from all of the possible resonant triads that involve the test wave identified by the point $(m,\omega)$.
For details on the derivation and on the r.h.s. of Eq.~\eqref{eq:1M} summarized above, we refer the reader to~\cite{iwthLPTN}.
%{\color{red}After integration over horizontal wavenumbers exploiting isotropy,}
The r.h.s. of Eq.~\eqref{eq:1M}  has the following form~\cite{LT2}:

{\color{black}
\begin{equation}\label{eq:8M}
    I[e,e](m_0,\omega_0) = J(m_0,\omega_0)
     \int  \Big[R^0_{12} - R^1_{02} - R^2_{10}\Big] d\mathbf p_{12}\,,
\end{equation}}
{\color{black}with $J(m,\omega)=4\pi\rho_0^2N^2m^2\omega^2/g^2$. The term $R^0_{12}$ reads

\begin{equation}
    R^0_{12} = 4\pi n_0 n_1n_2 \left(\frac{1}{n_0}-\frac{1}{n_1}-\frac{1}{n_2}\right) \vert V_{12}^0\vert ^2 \delta(\omega_0-\omega_1-\omega_2)\delta(\mathbf p_0 - \mathbf p_1  - \mathbf p_2)\,,
\end{equation}
with 3D action spectrum $n_{\mathbf p}=e(m,\omega)/J(m,\omega)$ and matrix elements

\begin{equation}\label{eq:2aa}
\begin{aligned}
&\left\vert V^0_{12}\right\vert^2 \\
&= \frac{N^2}{32g}\bigg\{\left[ \frac{\mathbf{k}_0 \cdot \mathbf{k}_1}{k_0 k_1}k_2 \frac{\omega_0\omega_1 {+} f_0^2}{\sqrt{\omega_0\omega_1\omega_2}}
+ \frac{\mathbf{k}_0\cdot \mathbf{k}_2}{k_0 k_2}k_1 \frac{\omega_0\omega_2 + f_0^2}{\sqrt{\omega_0\omega_1\omega_2}} +
 \frac{\mathbf{k}_1\cdot \mathbf{k}_2}{k_1 k_2}k_0 \frac{\omega_1\omega_2 - f_0^2}{\sqrt{\omega_0\omega_1\omega_2}} \right]^2\\
& \qquad +\left( \frac{f_0 \mathbf{k}_1\cdot\mathbf{k}_2^\perp}{k_0k_1k_2\sqrt{\omega_0\omega_1\omega_2}} \right)^2 \left[ \omega_0(k_1^2-k_2^2) + \omega_1(k_0^2-k_2^2) + \omega_2 (k_1^2-k_0^2) \right]^2\bigg\}\,.
\end{aligned}
\end{equation}
$\mathbf k = (k_x,k_y)$ denotes the 2D horizontal wavenumber, and $\mathbf k^\perp = (-k_y,k_x)$.}
A set of three wavenumbers 
simultaneously fulfilling the frequency and wavenumber delta functions is called resonant. The three resonance conditions in Eq.~\eqref{eq:8M} restrict the integration domain onto a 2D subset of $\mathbb R^6$ called the resonant manifold, with six independent branches -- two for each resonance condition. The six branches have analytical solution only in the non-rotating approximation ($f=0$)~\cite{dematteis2022origins}. 

Here, we depart from such an approximation and treat the fully rotating problem. {\color{black}For the numerical integration of the collision integral~\eqref{eq:8M}, we use the same discretization of the $k-m$ space as in~\cite{dematteis_lvov_2021}. The change of variables to the $m-\omega$ space is obtained by using the dispersion relation~\eqref{eq:disp}. The search for the solutions of the six branches of the resonant manifold (see Fig.~\ref{fig:2}) with $f\neq0$
is performed iteratively starting from the exact solutions of the $f=0$ case~\cite{dematteis_lvov_2021}, using the {\it fzero} {\it Matlab} routine.
This implies a significant increase in the computational cost compared to the non-rotating approximation.}

A representation of the six-branch resonant manifold for a test wave with coordinates $(m,\omega)$ is shown in Fig.~\ref{fig:7M}, as the test wave is moved around the $m-\omega$ space. For $\omega\gg f$, approaching the non-rotating limit, the resonant manifold tends to the analytical non-rotating solutions~\cite{dematteis2022origins}. As $\omega\to f$, on the contrary, the deformation of the resonant manifold becomes apparent, and only three branches are present for $\omega <2f$. A direct numerical integration over the two remaining degrees of freedom spanning the resonant manifold finally returns a numerical value for~\eqref{eq:8M}.

\subsection{Wave-wave interaction transfers}\label{sec:6.3.1a}

Given two non-overlapping subregions of the spectral space, sets ${A}$ and ${B}$, the instantaneous energy {\color{black}transfer} from ${A}$ (input control set) to ${B}$ (output control set) is given by

{\color{black}
\begin{equation}\label{eq:2}
\begin{aligned}
	P_{A\to B} &= -\frac{N^2}{g}\int_A   I[e,e] (m,\omega{ \vert B})  dm d\omega \,,\\
 I[e,e] (m_0,\omega_0{ \vert B})  
&= J(m_0,\omega_0)
     \int  \Big[\Big(\frac{\chi_B(\mathbf p_1)\omega_1 + \chi_B(\mathbf p_2)\omega_2}{\omega_1+\omega_2}\Big)R^0_{12} \\
     & \quad\qquad\qquad\qquad - \chi_B(\mathbf p_1) R^1_{02} - \chi_B(\mathbf p_2) R^2_{10}\Big] d\mathbf p_{12}
 \end{aligned}
\end{equation}}
where $I[e,e] (m,\omega { \vert B})$ captures the resonant transfers of mode $(m,\omega)\in {A}$ subject to the constraint that the
transfer must be between point $(m,\omega)$ and set $B$~\cite{dematteis2023structure}. {\color{black}The indicator function of set $B$ is defined so that $\chi_B(\mathbf p)=1$ if $\mathbf p\in B$ and $\chi_B(\mathbf p)=0$ otherwise. The factor $N^2/g$ takes into account the conversion between volumes from isopycnal coordinates to Eulerian coordinates, so that the energy transfers are expressed in W/kg.}

In order to go from Eq.~\eqref{eq:1M} to \eqref{eq:2}, the following two steps are necessary: (i) A logical weighting that accepts only those contributions whose output wavenumbers are in the output set $B$, discarding the rest of the contributions; (ii) An outer supplementary 2D integration over the input set ${A}$.
The method is rigorously derived in~\cite{dematteis2023structure}. Here, the main innovation is the use of a $3\times3$ multiscale partition of the Fourier space tailored to the oceanic internal-wave problem. A complete knowledge of the inter-scale transfers requires a permutation of all of the nine subregions of the partition in the roles of ${A}$ and ${B}$, resulting into a $9\times9$ energy transfer matrix -- although not all elements are relevant to the analysis: e.g., we are not interested in quantifying transfers between different dissipative subregions. The matrix is anti-symmetric, satisfying some fundamental properties for a directed transfer: (i) $ P_{A\to A}=0\,$, (ii) $ P_{A\to B} = - P_{B\to A}\,$, and (iii) $ P_{A\cup B\to C}=  P_{A\to C} +  P_{B\to C}$ {\color{black}(for $A\cup B = \emptyset$)}.

{\color{black}We end this section with a remark on numerical details.  We use the antisymmetry of $ P_{A\to B}$ as a criterion to establish a satisfactory numerical resolution level of the spectral space, in the numerical integration. Our convergence requirement is that $( P_{A\to B}+ P_{B\to A})/ P_{A\to B}$ (our measure of numerical uncertainty, as the quantity tends to zero as the resolution increases) be smaller than $5\%$ for all elements of the transfer matrix.} 
{\color{black}Using parallel computing on a 12-core station, the computation of the $9\times9$ transfer matrix for a given point in parameter space $(s_{\rm NI},s_\omega,s_m)$ takes about 6 hours. We use a numerical grid of $5\times5\times5$ points to discretize the parameter space, for a total computing time of the order of a month. Once this computation has been performed once and stored in memory, we then use 3D interpolation to establish the flux matrix associated with any given point $(s_{\rm NI},s_\omega,s_m)$ that is not on the numerical grid, with almost negligible extra computing time. The processing of the hundreds of thousands of spectra in the combined global data \textcolor{black}{set}, %(results in section~\ref{sec4b})
including both spectral slope fitting and energy transfer predictions, is performed in a few hours of computing time. All computations and plotting in the present manuscript are performed in {\it Matlab}.}

\subsection{First-principle mixing parameterization}\label{sec:6.3.2}
The turbulent energy production rate, $\mathcal P$, is given by the sum of the energy transfers across the boundary delimiting the dissipative regions~\cite{mccomas1981dynamic,eden2019numerical}. The resulting first-principles formula for $\mathcal P$ reads:

\begin{equation}\label{eq:3}
	\mathcal P = \mathcal P_0^{\rm th}(s_{\rm NI},s_\omega,s_m) \frac{f}{f_0}\left(\frac{E{m_*}^{s_m-1}}{E_0{m_{*}^0}^{s_m-1}}\right)^2\,,
\end{equation}
where $\mathcal P_0^{\rm th}(s_{\rm NI},s_\omega,s_m)$ is a function in the 3D space of the spectral slopes.

Let us summarize the main steps leading from the primitive equations of the ocean to the formula~\eqref{eq:3} for $\mathcal P$.
\begin{itemize}
    \item Derivation of Eq.~\eqref{eq:1M} under the assumptions listed in section~\ref{sec:6.3.1};
    \item Calculation of the inter-scale energy transfers between the different subregions of the Fourier space (section~\ref{sec:6.3.1a}), via Eq.~\eqref{eq:2}, {\color{black} including the total energy transfer into the dissipative subregions, leading directly to formula~\eqref{eq:3}}.
    %\item Calculation of the total energy transfer into the dissipation scales as the sum of all inter-scale transfers going across the boundary with the dissipative subregions, leading directly to formula~\eqref{eq:3}.
\end{itemize}

The function $\mathcal P_0^{\rm th}(s_{\rm NI},s_\omega,s_m)$ is computed numerically and is represented in Fig.~\ref{fig:6M} as a colormap at five given values of $s_m$.
$\mathcal P_0^{\rm th}$ is independent of $m_*$ because of the following approximation. As shown in Fig.~\ref{fig:2} the direct contribution to $\mathcal P$ from small wavenumbers close to $m_*$ is nearly negligible. By the analytical properties of the GM76 spectrum~\cite{regional}, {\color{black}for fixed slope $s_m$ and total energy $E$, the vertical wavenumber spectrum is given by 

\begin{equation}\label{eq:17}
    e(m)=\frac{c(s_m)E(m_*)^{-1}}{1+\left(\frac{m}{m_*}\right)^{s_m}},\qquad \text{with}\quad  c(s_m) = \frac{s_m}{\Gamma\left(\frac{1}{s_m}\right)\Gamma\left(\frac{s_m-1}{s_m}\right)},
\end{equation}
where $\Gamma(\cdot)$ is the gamma function (see section 2.2 of \cite{regional}). At high wavenumbers, we have $e(m)\simeq Ec(s_m)(m_*)^{s_m-1}m^{-s_m}$, for $m\gg m_*$. This shows that the high-wavenumber spectral level depends directly on the value of $m_*$, even for $m\gg m_*$. Thus, in order for the spectrum with the reference parameter $m_*^0$ to match the high-wavenumber spectral level of a spectrum with parameter $m_*$, it has to be renormalized by a factor $(m_*/m^0_*)^{s_m-1}$. This explains the correction to the energy level in Eq.~\eqref{eq:3}.}

{\color{black} Here, we briefly derive a relationship between the normalized shear spectral level of Eq.~\eqref{eq:3}~\cite{polzin2014finescale} and our five spectral parameters. Using Eq.~\eqref{eq:17} and the definition of shear spectral density $S_z(m)=2m^2e_k(m)$, where $e_k(m)=e(m)R_\omega/(1+R_\omega)$, we obtain the high wavenumber approximation

\begin{equation}\label{eq:18}
    S_z(m) \simeq 2\frac{R_\omega}{1+R_\omega} c(s_m) E m_*^{s_m-1} m_c^{3-s_m}.
\end{equation}
Following~\cite{polzin2014finescale}, only in this paragraph we use $m_c$ as the high wavenumber cutoff defined via

\begin{equation}\label{eq:19}
    \int_0^{m_c} S_z(m) dm =\frac{2\pi}{10}N_0^2\,.
\end{equation}
After some algebra, using the definition of the normalized shear spectral level as $ {\mathcal E} = 0.1$cpm$/m_c$, where cpm $=2\pi/$metre, we arrive at

\begin{equation}\label{eq:20}
     {\mathcal E} = \left(\frac{2\pi}{10}N_0^2 \frac{1}{2c(s_m)}\frac{1+R_\omega}{R_\omega} \frac{3-s_m}{Em_*^{s_m-1}}\right)^{\frac{1}{s_m-3}}\,.
\end{equation}
When $s_m=2$, $\mathcal E\propto Em_*$. As a proof of consistency, if we consider the case of the standard GM76 reference parameters (which give $c(s_m)=2/\pi$ and $R_\omega=3$) we correctly obtain $\mathcal E=1$.
}

%Notably, the factor of $E m_*^{s_m-1}$ consequently appearing in the formula~\eqref{eq:3} is proportional to the shear-variance density, making an important direct connection with the original FP formula~\eqref{eq:0}~\cite{polzin1995finescale}.
%Add here clarification abour shear spectrum.

%, and in the numerical scripts in the Supplementary Material.

In steady conditions, the {\color{black}turbulent energy production rate} feeds the vertical buoyancy fluxes, $\mathcal B$, and the dissipation into disordered molecular motion, $\epsilon$: $\mathcal P = \mathcal B + \epsilon$. \textcolor{black}{The vertical mixing diffusivity $K$ is defined as a function of the buoyancy fluxes via $K = \mathcal B/N^2$}. We use the classical Osborn parameterization~\cite{osborn1980estimates,thorpe04}, establishing the proportion to be $\mathcal B=R_f \mathcal P$, where $R_f\simeq0.17$ is the flux Richardson number~\cite{garabato2022oceanBook}. Thus, diapycnal diffusivity and turbulent dissipation rate are estimated, respectively by

\begin{equation}
\label{eq:4}
    K=\frac{R_f \mathcal P}{N^2}\,,\qquad \epsilon = (1-R_f)\mathcal P\,.
\end{equation}

The five-parameter spectral simplification is a powerful conceptualization that turns out crucial for the realizability of our analysis in terms of numerical cost. % (details in section~\ref{sec:6.4}). %allows us to perform the heavy numerical calculation of $\mathcal P_0(s_{\rm NI},s_\omega,s_m)$ upfront. Once it has been computed (requiring some weeks of computing time), we are then able to process thousands of different spectra and obtain the respective theoretical mixing estimates efficiently (a few hours of computing time). Details on the numerical computation are provided in section~\ref{sec:6.4}
However, in principles any particular observed 2D spectrum can be processed individually without reduction to the five-parameter approximation.
{\color{black}In conclusion, we can state that the calculation of $\mathcal P$ from a given input spectrum is from first principles. Yet, the input from observational spectra is phenomenological due to the fitting procedure to formula \eqref{eq:en-spec}. Indeed, also the use of~\eqref{eq:4} to pass from $\mathcal{P}$ to the observables $K$ and $\epsilon$ is empirical. The problem of mixing efficiency, i.e. the ratio of $K$ and $\epsilon$ in stratified turbulence, is somewhat decoupled from the theoretical investigations of the present manuscript. On its own, it defines an entire area of intense research activity~\cite{peltier2003mixing,bluteau2013turbulent,maffioli2016mixing,ijichi2018observed,gregg2018mixing,ijichi2020variable}.}

%where $\rho\simeq10^3$ kg$\,$m$^{-3}$ and  $R_f\simeq0.17$~\cite{osborn1980estimates,thorpe04}.

%Remarkably, the phenomenological FP estimate for the GM76 spectrum in Fig.~\ref{fig:2}, of $\mathcal P=8\times10^{-10}$ W kg$^{-1}$~\cite{polzin2014finescale}, is retrieved quite closely by our theoretical prediction of $\mathcal P=9.8\times10^{-10}$ W kg$^{-1}$.

\subsection{Interaction processes: classification and quantification}
\label{sec:6.3.3}

%{\color{red}Condense Figs. 7 and 8 into a single figure with only half of the panels (redundant details)}

\begin{figure*}[ht]%
\centering
\includegraphics[width=\textwidth]{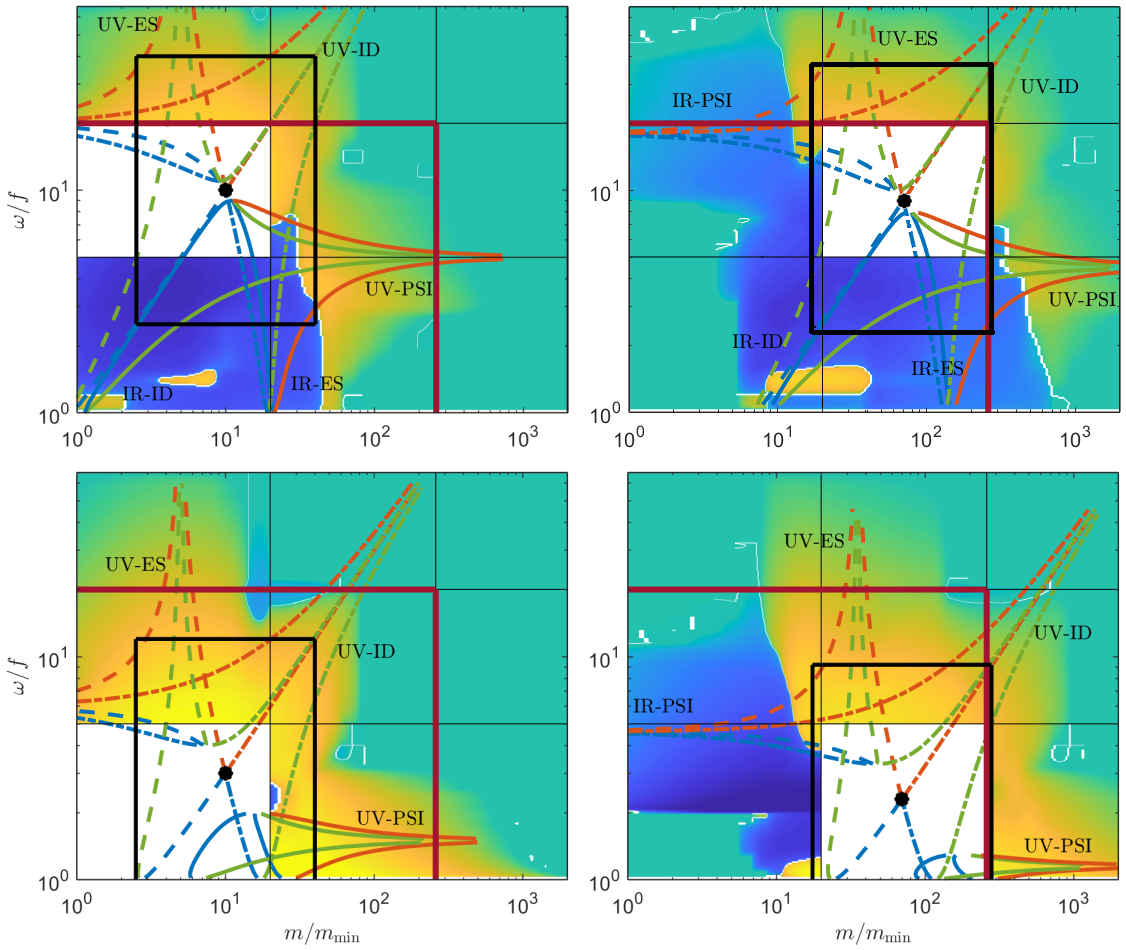}
\caption{{\bf Evaluation of energy transfers represented in spectral space.} Contribution $I[e,e] (m,\omega { \vert B})$, representing the energy exchange between a point $(m,\omega)$ and set $ {B}$, the white region in each of the four panels. A yellow tonality at a point $(m,\omega)$ indicates energy directed from $ {B}$ to $(m,\omega)$, and a blue tonality the reverse. Green points have no direct energy exchange with $ {B}$. The curved lines are the resonant manifold for a representative point (black dot) in set $ {B}$. The matching between the resonant manifold's lobes and the observed energy transfer patterns allows us to discern the direction of the energy transfers associated with the different processes of Induced Diffusion (ID), Parametric Subharmonic Instability (PSI), and Elastic Scattering (ES). The ultraviolet (UV) and infrared (IR) labels indicate interaction of the test wave with waves of smaller and larger scale, respectively. {\color{black}The black box surrounding the test wave delimits a scale separation by a factor of 4 from the test wave, delimiting the scale-separated interactions from the spectrally-local interactions (see Fig.~\ref{fig:2}E).}}\label{fig:7M}
\end{figure*}

An in-depth analysis of the contributions to the energy transfers allows us to identify the transfers due to each of the resonant branches with a known physical process. The numerical procedure underlying this analysis is depicted in  Fig.~\ref{fig:7M} for the different spectral subregions and the different interaction classes. %More details are given in Supplementary Information. 
The resonant manifold associated with a given test wave $(m,\omega)$ (shown in Fig.~\ref{fig:2}E and Fig.~\ref{fig:7M}) has six lobes corresponding to the six resonant branches. Asymptotically, i.e. far away from the test wave, these lobes identify the three well-known scale-separated regimes of internal wave-wave interactions: ID, PSI, and ES~\cite{McComas1977,mccomas1981dynamic,Muller86}. Each of the three processes corresponds to two lobes: one relating the test wave to larger scales (termed ultraviolet, or UV), and the other relating the test wave to smaller scales (termed infrared, or IR). We remark that our use of the terms ID, PSI, and ES refers to the whole resonant lobes, that technically correspond to the original definition of the three classes of~\cite{McComas1977} only in the asymptotic scale-separated regimes.

% HELL0 FROM THE GULF STREAM  :) (ARNAUD)
% LOL HI ARNAUD, greatings to the fearless sailor!
In Fig.~\ref{fig:7M}, we represent the contribution $ I[e,e] (m,\omega  \vert {B})$ (cf. Eq.~\eqref{eq:2})) for a reference spectrum~\cite{GM76}, choosing the four non-dissipative subregions as set ${B}$, respectively in the four panels. The yellow color indicates wavenumbers that are gaining energy from set ${B}$ (white box), while the blue color indicates wavenumbers that are transferring energy into set ${B}$. Clearly, three lobes toward smaller scales and three lobes toward larger scales (when applicable) appear in each of the panels. We superpose the curves of the resonant manifold for a representative test wave in each box to illustrate that the directions of the observed energy transfer correspond to the three main physical mechanisms of ID, PSI, and ES. Moreover, the energy transfer direction is {\color{black}predominantly} from the IR regions (blue) to the UV regions (yellow), i.e. from the large scales to the small scales. {\color{black}In summary, in Fig.~\ref{fig:7M} each of the three processes is associated with a characteristic direction of energy propagation: the ``energy cascade'' is direct in both wavenumber and frequency for the ID branch, direct in wavenumber and slightly inverse in frequency for the PSI branch, slightly inverse in wavenumber and direct in frequency for the ES branch.}  %section~\ref{sec3}.

\section*{Data availability} The analyses of the Argo data used in this manuscript are publicly available at \url{https://doi.org/10.5281/zenodo.6966416} and \url{ https://doi.org/10.17882/95327}. Details on the availability of the current-meter data are found in~\cite{le2021variability} -- please contact aleboyer@ucsd.edu for further information. Details on the High-Resolution Profiler data are found in~\cite{polzin1995finescale,regional}.
\section*{Code availability} The codes developed for the data analysis and theoretical computations in this manuscript are available at \url{https://doi.org/10.5281/zenodo.12529645}~\cite{dematteis2024}.

\pagebreak

\section*{Supplementary information}

\subsection*{Analysis of nonlinearity level}\label{sec:6.3.4}

\subsubsection*{Coarse-grained and pointwise nonlinearity levels}

We define two distinct (coarse-grained) nonlinear times. The first, $\tau_{\rm nl}$, is the timescale of nonlinear evolution of the wave energy spectrum; this is obtained as the ratio between the total energy in a given subregion of spectral space, and the time rate of change of energy in the subregion. The second, $\tau_{\rm res}$, is the energy residence time in the subregion, given by the ratio of the total energy in the subregion, and the total energy flux transferred out of the subregion -- i.e., the ``sum'' of the red arrows going out of the rectangle representing a given subregion in Fig.~2 in Main Text. As was argued by~\cite{holloway1980oceanic}, it is the second, the residence time $\tau_{\rm res}$, to be the relevant time for the energy transfers, although the definition of~\cite{holloway1980oceanic} is pointwise in spectral space -- see Eq. (1) below. It is sufficient to think of a nonequilibrium stationary state with a nonzero flux to realize that in such case, since the spectrum does not change over time, we have $\tau_{\rm nl}=\infty$, but $\tau_{\rm res}$ will have a finite value that depends on the intensity of the flux. For this reason, we opt to base our analysis of nonlinearity on $\tau_{\rm res}$.

\begin{figure*}[h]%
\centering
\includegraphics[width=.8\textwidth]{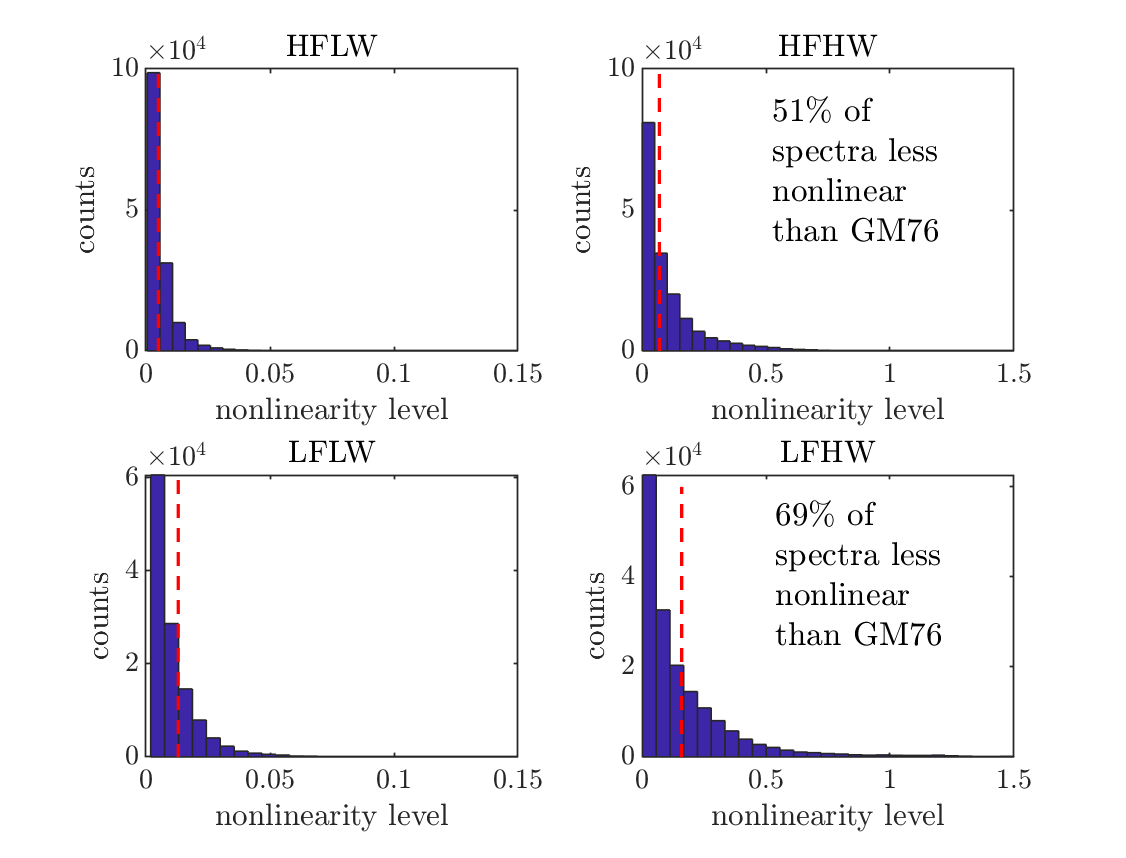}
\caption{Histograms of coarse-grained nonlinearity level in the same scale ranges as Fig.~3{\bf D} in Main Text.  %(section~\ref{sec4d})
The HFHW scales have less than $1\%$ of the occurrences at $r_{\rm nl}>0.5$. The LFHW scales have about $2\%$ of the area at $r_{\rm nl}>1$ and about $7\%$ of the area is at $r_{\rm nl}>0.5$. {\color{black}For these coarse-grained metrics, a value smaller than 1 does not ensure weak nonlinearity, since the pointwise metric may be small in a subregion, but well above 1 in another subregion ({\it cf.} Fig. S2)}. The coarse-grained nonlinearity level for the GM76 spectrum is indicated by the  dashed red lines.}\label{fig:4}
\end{figure*}

We now compare the estimated residence times with the linear period of internal waves, {\color{black}$\tau_{\rm lin}(\omega)=2\pi/\omega$}, a classical way to assess the nonlinearity strength compared to linear dispersion~\cite{Muller86}. % For the weakly nonlinear theory of section~\ref{sec3} to be valid, the nonlinear times must be much slower than the linear times. 
In terms of the nondimensional nonlinearity level $r_{\rm nl} = \tau_{\rm lin}/\tau_{\rm res}$, a weakly nonlinear regime is defined by $r_{\rm nl}\ll 1$.
In Fig.~\ref{fig:4}, we show the distributions of $r_{\rm nl}$ in the combined global dataset. At low wavenumber, nonlinearity is always extremely weak. For low wavenumber and high frequency, $r_{\rm nl}$ remains smaller than 0.5, whereas, for high wavenumber and low frequency, the most nonlinear scales, about $2\%$ of the occurrences have $r_{\rm nl}>1$.

\begin{figure*}[h]%
\centering
\includegraphics[width=.97\textwidth]{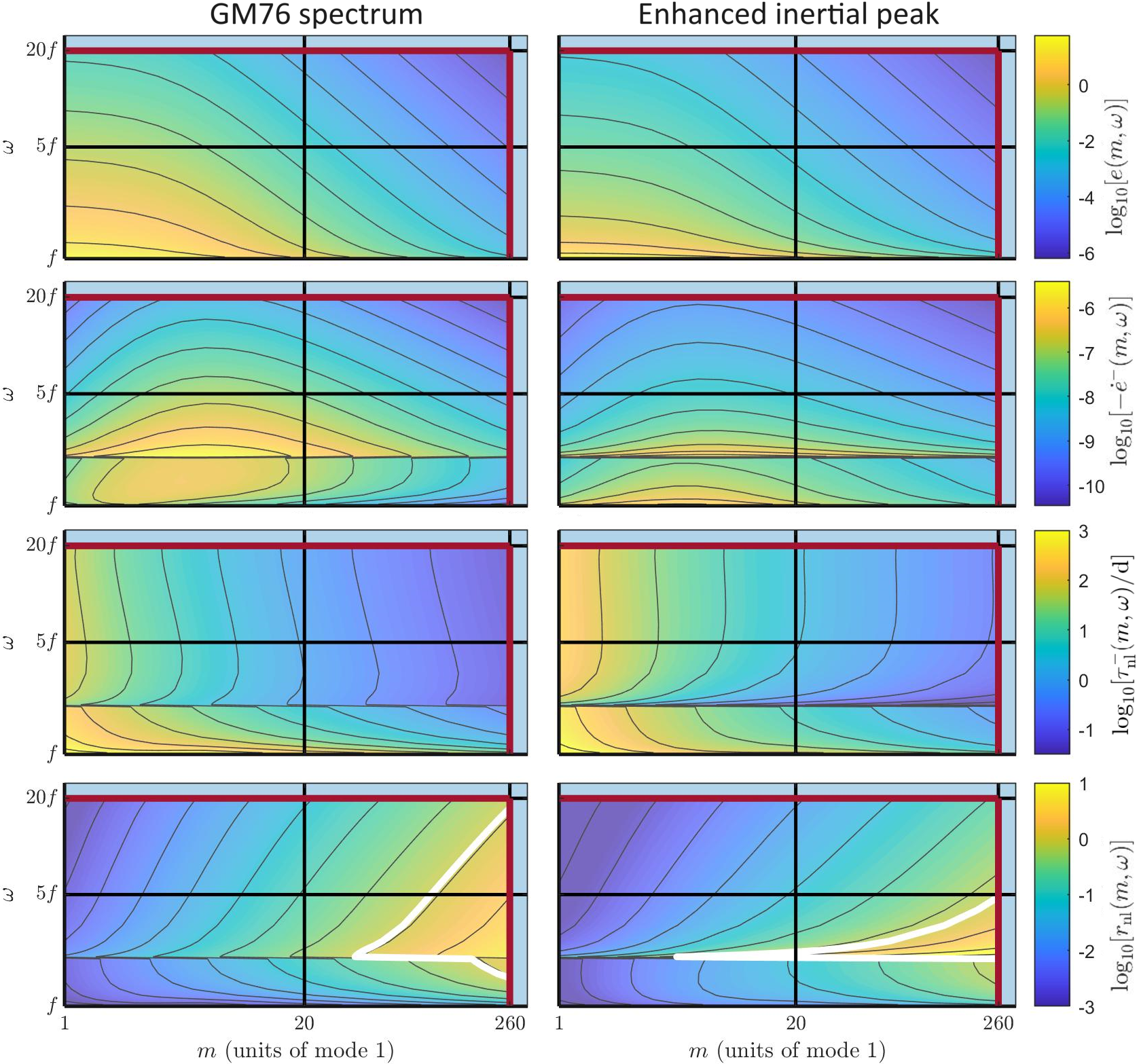}
\caption{Left and right panels are for the same two representative spectra shown in Fig.~2 (GM76 and similar spectrum with enhanced near-inertial peak), respectively. From top to bottom, we show color maps of the quantities necessary to evaluate the point-wise nonlinear residence time and nonlinearity level. In the color maps of $\tau_{\rm nl}^-(m,\omega)$ we observe time scales that go from hundreds of days for the near-inertial, low-wavenumber waves, up to a few hours for the high wavenumber waves. In the bottom panels, the white contour line is at $r_{\rm nl}(m,\omega)=1$. To the right of this line, nonlinearity dominates over dispersion, threatening the applicability of weak wave-wave interaction theory. The extra vertical red line in these panels is at $m=m_c/2$ (where $m_c$ is given by Eq. (3) in the main text), the value used in the sensitivity test depicted in Fig.~\ref{fig:9M} to reduce the amount of strongly nonlinear scales.}\label{fig:8Mb}
\end{figure*}

We now turn to a more detailed, pointwise generalization of the residence time, similar to that of~\cite{holloway1980oceanic}, illustrated for two reference spectra.
The pointwise residence time $\tau_{\rm nl}^-(m,\omega)$ is defined as the ratio between the energy density $e(m,\omega)$ and the negative-signed contributions to the rate of change of energy density, $\dot e^-(m,\omega)$:

\begin{equation}
    \tau_{\rm nl}^-(m,\omega) = \frac{e(m,\omega)}{\vert\dot e^-(m,\omega)\vert}\,.
\end{equation}
Again, in order to analyze the weakness of nonlinearity compared to linear dispersion, we compare the residence time with the linear wave period {\color{black}$\tau_{\rm lin}(\omega)$}. The pointwise quantities here defined and the nonlinearity parameter $r_{\rm nl}$, defined as

\begin{equation}
    r_{\rm nl}(m,\omega) = \tau_{\rm lin}(m,\omega)/\tau_{\rm nl}^-(m,\omega)\,,
\end{equation}
are shown for the two chosen reference spectra in Fig.~\ref{fig:8Mb}.
We observe a sharp discontinuity at $\omega=2f$. This is due to the disappearance of the PSI transfer mechanism -- very efficient for frequencies just larger than $2f$ -- at frequencies smaller than $2f$. The most nonlinear scales are located in proximity of $\omega=2f$, and toward the large wavenumbers. %, confirming the coarse-grained results of section~\ref{sec4d}.

{\color{black}Clearly, values of the coarse-grained $\tau_{\rm res}$ smaller than unity in a spectral region may still correspond to values larger than unity for the point-wise $\tau_{\rm nl}^-$ in some smaller spectral subregion (see Fig. S2). However, the coarse-grained metrics in Fig.~\ref{fig:4} are important for a relative comparison among different spectra, noting that we are not able to show a pointwise analysis for the large number of observed spectra. The coarse-grained nonlinearity levels for the GM76 spectrum are shown as dashed red lines in Fig.~\ref{fig:4}. In the most nonlinear spectral region (low frequency and high wavenumber), about $69\%$ of observed spectra are less nonlinear than GM76, and $31\%$ are more nonlinear than GM76.}

\subsubsection*{Comparison with Holloway (1980)}

\begin{figure*}[h]%
\centering
\includegraphics[width=\textwidth]{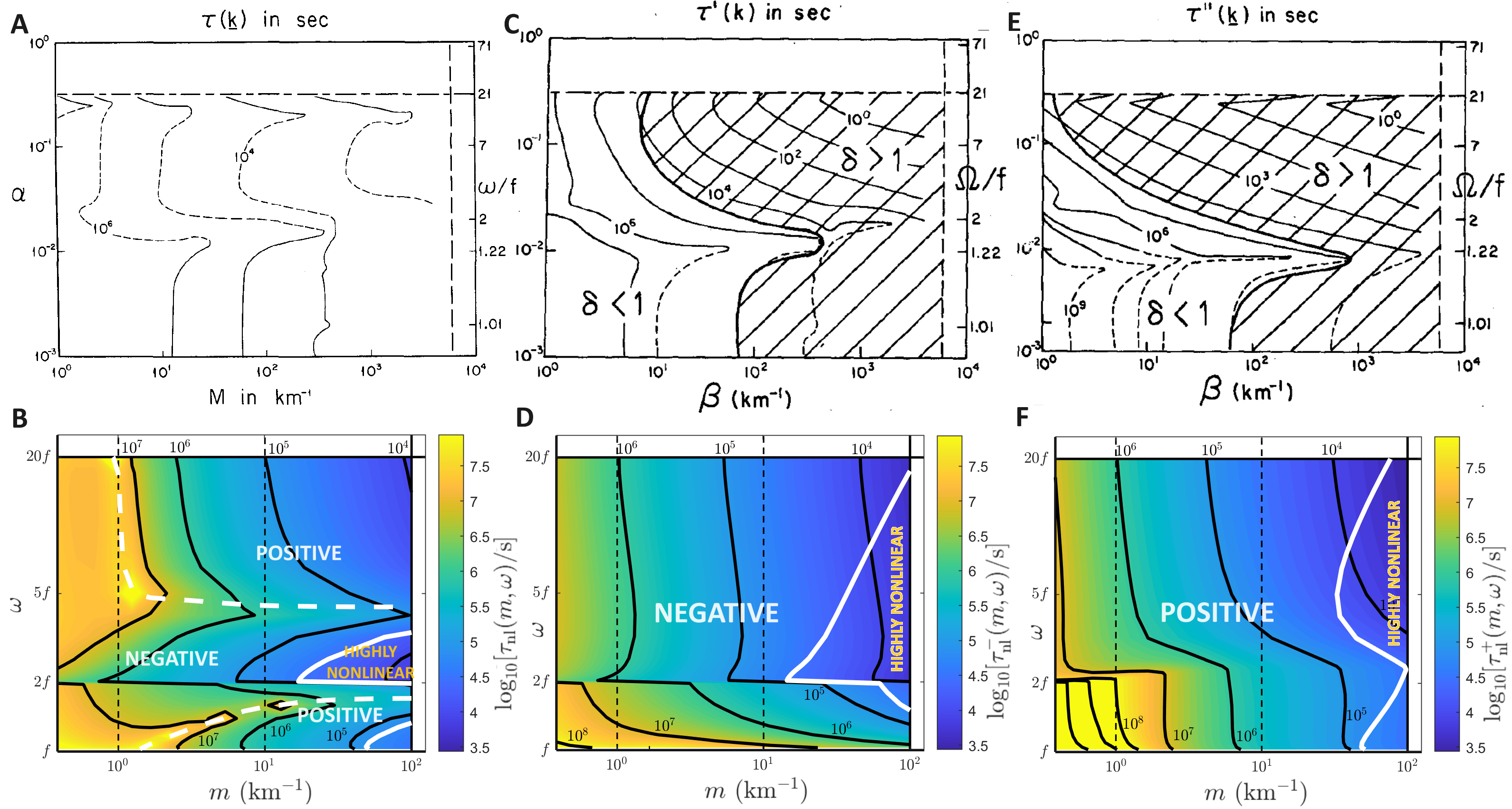}
\caption{\color{black}Detailed comparison between the plots in Holloway (1980)~\cite{holloway1980oceanic}, taken from~\cite{mccomas1977equilibrium} (top), and our results (bottom). The top panels show the spectral evolution timescale $\tau_{\rm nl}$. The negative contributions in {\bf A} (dashed contour lines) extend from below $2f$ up to $15f$, rather than from below $2f$ up to $5f$. The two plots differ for frequency larger than $2f$-$3f$. {\bf C} and {\bf D} should show comparable quantities (residence times), as argued in the text, but they clearly differ in the upper portion of the spectral space. This leads~\cite{holloway1980oceanic} to consider most of the high-frequency region highly nonlinear, whereas we find high nonlinearity at high wavenumber, but not at high frequency. {\bf E} (symmetrization timescale) and {\bf F} (re-filling timescale of an initially depleted mode) do not compare directly, although they show some similarities. Again, though, they lead to drastically different assessments of high-nonlinearity regions. Note that the high-nonlinearity regions in panels {\bf D} and {\bf F} are superposed effects with opposite sign, and the resulting evolution timescale ({\bf B}) is shorter than the nonlinear time in a much smaller region. It is the latter (inverse of the so-called Boltzmann rate) that is usually taken in consideration as a parameter for high nonlinearity in the weak wave turbulence literature \cite{NazBook,regional}.}\label{fig:9Ma}
\end{figure*}

{\color{black}Here, we discuss the results of~\cite{holloway1980oceanic}, where oceanic internal waves were (quite categorically) declared too nonlinear to be described by weakly-nonlinear wave-wave interactions. First, the conclusions of~\cite{holloway1980oceanic}, based on results by~\cite{mccomas1977equilibrium}, reached conclusions that reverberate to the present days based on the energy level of the GM76 spectrum alone. Note that the broad distribution of energy levels in the combined global dataset presented in our work is such that more than two thirds of spectra are less nonlinear than GM76 in the most nonlinear range of scales (Fig.~\ref{fig:4}). Second, our numerical results differ from those of~\cite{mccomas1977equilibrium}. We show this in detail in Fig.~\ref{fig:9Ma}. The top panels represent the same quantity $\tau_{nl}(m,\omega)=e(m,\omega)/\dot e(m,\omega)$ in~\cite{mccomas1977equilibrium} ({\bf A}) and in our calculation ({\bf B}). The hydrostatic-balance assumption implemented in our code is well satisfied for the considered range of frequencies up to less than $N/3$, and our result is consistent with recent independent calculations~\cite{regional,eden2019numerical,wu2023energy} (no matter whether near-resonant or exactly resonant, hydrostatic or non-hydrostatic). All of these recent results support our assessment that the time rate of change of energy is negative in a frequency band roughly between $2f$ and $4f$--$5f$. More closely, our result recovers exactly the pattern obtained in~\cite{regional} in a near-resonant setup. On the contrary, the results of~\cite{mccomas1977equilibrium} show a negative time rate of change of energy up to frequencies larger than $10f$, where a sharp change of sign departs very markedly from our results. While we are not able to check the original code of~\cite{mccomas1977equilibrium}, we attribute such change to the abrupt numerical cutoff that apparently was implemented for all frequencies larger $21f$. We suggest that this cutoff is responsible for the abrupt nonphysical upper change of sign -- on the other hand, the lower change of sign in proximity of $2f$ is well understood as due to the onset of the PSI branch. We conclude that the results of~\cite{mccomas1977equilibrium} are not reliable for frequencies larger than about $3f$ (while they are roughly in agreement with our results for lower frequency).}

{\color{black}We did not reproduce the same experiment of decay of a narrow perturbation as in~\cite{mccomas1977equilibrium} (Fig.~\ref{fig:9Ma}{\bf C}), but we can associate to this decay timescale the pointwise residence time $\tau_{\rm nl}^-$ (Fig.~\ref{fig:9Ma}{\bf D}), as discussed in the previous section. Again, a clear departure between the two plots is observed at frequencies larger than $2f$. In particular, the ``linear time equals nonlinear time'' line drawn by~\cite{holloway1980oceanic} on top of the plot by~\cite{mccomas1977equilibrium} does not have correspondence in our results. The analogue in Fig.~\ref{fig:9Ma}{\bf D} is the white line on the right side of the plot.}

{\color{black}Finally, we report in Fig.~\ref{fig:9Ma}{\bf E} the experiment by~\cite{mccomas1977equilibrium} about the symmetrization timescale of an initially asymmetric spectrum, which shows closer qualitative similarity with the timescale $\tau_{nl}^+(m,\omega)=e(m,\omega)/\dot e^+(m,\omega)$ in our analysis, where $\dot e^+(m,\omega)$ denotes the positive-signed contributions to the rate of change of energy
density. $\tau_{nl}^+$ is the timescale it takes a wavenumber to restore its initially depleted energy density, re-equilibrating with the rest of the spectrum. Even for this different timescale, the high-nonlinearity levels concentrate on the far right end of the plot, contrary to the line drawn by ~\cite{holloway1980oceanic}. In light of our results, and of the quantitative support they find in independent recent analyses~\cite{regional,eden2019numerical,wu2023energy}, we consider the quantitative conclusions by~\cite{holloway1980oceanic} unreliable.} %To say the least, the categorical message of~\cite{holloway1980oceanic} should be severely toned down, since oceanic internal waves are, in fact, often weakly nonlinear in most of their scales.

{\color{black}For a GM76 spectrum, Fig.~\ref{fig:9Ma} shows that when the nonlinear timescales at play become comparable with the linear time, this happens for wavenumbers that are close to the $m_c=2\pi/10$ m$^{-1}$ boundary. Moreover, cancellations between positive and negative contributions to the time rate of change $\dot e(m,\omega)$ are such that the spectral evolution timescale $\tau_{\rm nl}(m,\omega)$ shown in Fig.~\ref{fig:9Ma}{\bf B} is shorter than the linear time only in the restricted range of scales in which $\dot e(m,\omega)<0$. In weak wave turbulence it is this timescale $\tau_{\rm nl}(m,\omega)$ (inverse of the so-called Boltzmann rate~\cite{regional}) that is usually considered as a proxi for nonlinearity~\cite{NazBook}: comparable positive and negative contributions to $\dot e(m,\omega)$ (Figs.~\ref{fig:9Ma}{\bf D} and~\ref{fig:9Ma}{\bf F}, respectively) will coexist and counter each other decreasing the effective nonlinearity (Fig.~\ref{fig:9Ma}{\bf B}).}

{\color{black}Moreover, the small nonlinear times $\tau_{\rm nl}(m,\omega)$ in Fig.~\ref{fig:9Ma}{\bf B} are an indication of the nonstationary conditions represented by the GM76 spectrum. If evolved forward in time, this would lead to a depletion of energy density in the frequencies just larger than $2f$ and at large wavenumber, and to a spectral distribution that is not separable in the frequency-wavenumber space. The observational trace of this specific issue in the form of a ``mid-frequency dip'' was discussed in section 5.1.2.3 of~\cite{regional}. In this manuscript, our global-scale approach forces us to necessarily assume separability of spectra in frequency and wavenumber, so that we can approximate 2D spectra by cross-matching independent databases. Considering the time evolution would require high computational cost, and a detailed knowledge of the energy sources of the internal wavefield and of actual 2D spectra -- a type of information that is rarely simultaneously available~\cite{regional}. Such studies, which are the subject of ongoing research, require focusing on few select locations rather than on global patterns, outside the scope of the current work.}

{\color{black}The large number of spectra processed in our analysis requires us to assume a large wavenumber scale $m_c$ past which to evaluate the energy transfers. We made the choice of $10$ m for this vertical threshold scale, due to its large significance in the literature~\cite{regional,eden2019numerical,garabato2022oceanBook}. We are aware that dealing rigorously with the issue of high nonlinearity would imply a choice of $m_c$ that is case dependent. This would compromise the feasibility of global-scale evaluations such as the ones that we perform.}

{\color{black}Here, our justification of this approach has a pragmatic basis. Below, a sensitivity test shows that our evaluations are weakly sensitive to the choice of the $m_c$ threshold. This near-independence of the threshold choice makes our results effectively free of any arbitrary degree of freedom.
Therefore, we chose to keep in our analysis spectra that have an elevated energy level and should raise a high-nonlinearity warning. A few considerations are in order: First, in absence of any alternative prognostic fully-nonlinear theories~\cite{whalen2020internal}, the weakly nonlinear prediction is the only theoretical explanation available that does not include adjustable parameters. Second, when averaging over time and among different regions with similar turbulence production levels, the agreement between our first-principles weakly nonlinear predictions and the phenomenological strain-based FP predictions is striking (see Fig. 3{\bf B} in main text). If anything, this is a clear {\it a posteriori} indication that weakly nonlinear interactions underly internal-wave driven turbulent mixing in the global ocean interior. Characterizing location-dependent departures due to violation of weak nonlinearity, but also possibly of other assumptions, are outside the scope of this manuscript, and are the subject of ongoing research.} %E.g., departure from weakly nonlinear predictions in a highly energetic submarine canyon are currently being undertaken (cite Charlotte if available before resubmission).

\subsection*{Sensitivity analyses}
\subsubsection*{High-wavenumber boundary}\label{sec:6.4.1}

\begin{figure*}[h]%
\centering
\includegraphics[width=.5\textwidth]{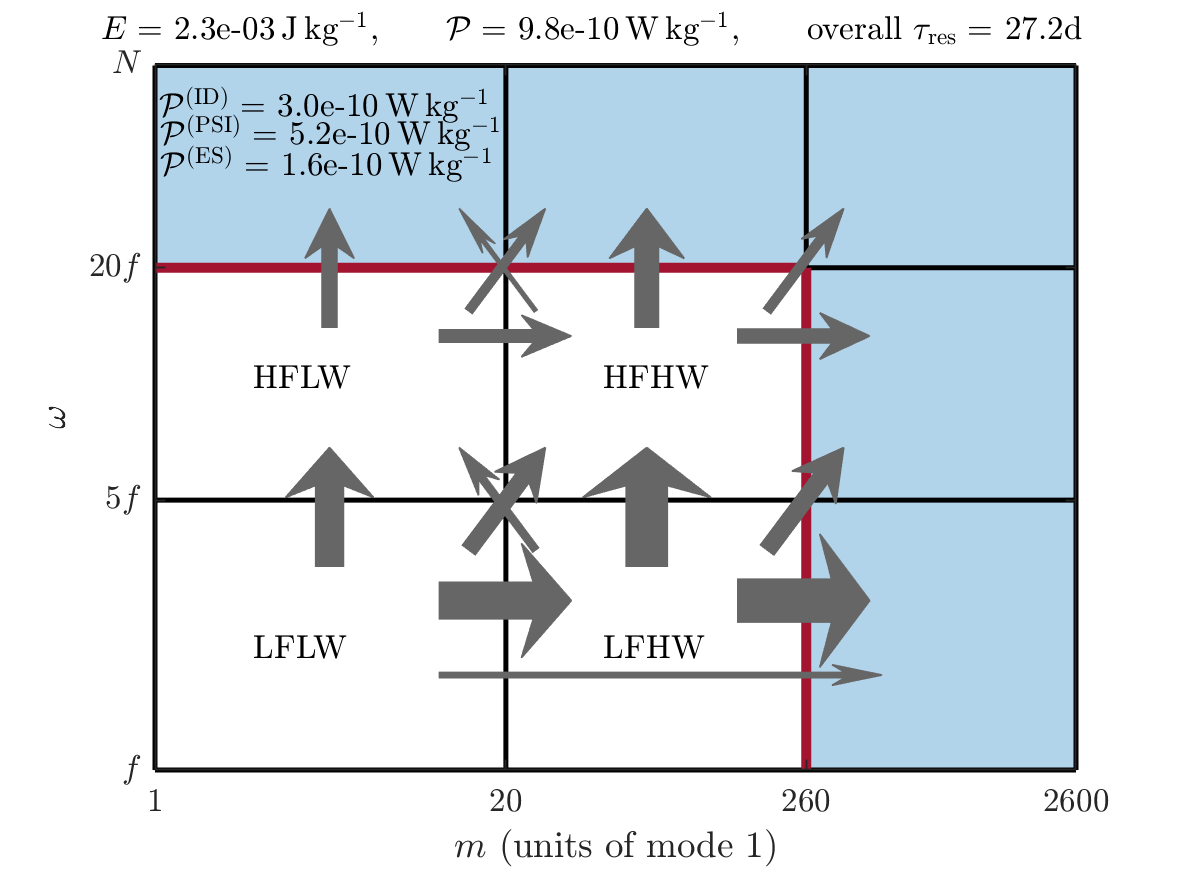}%
\includegraphics[width=.5\textwidth]{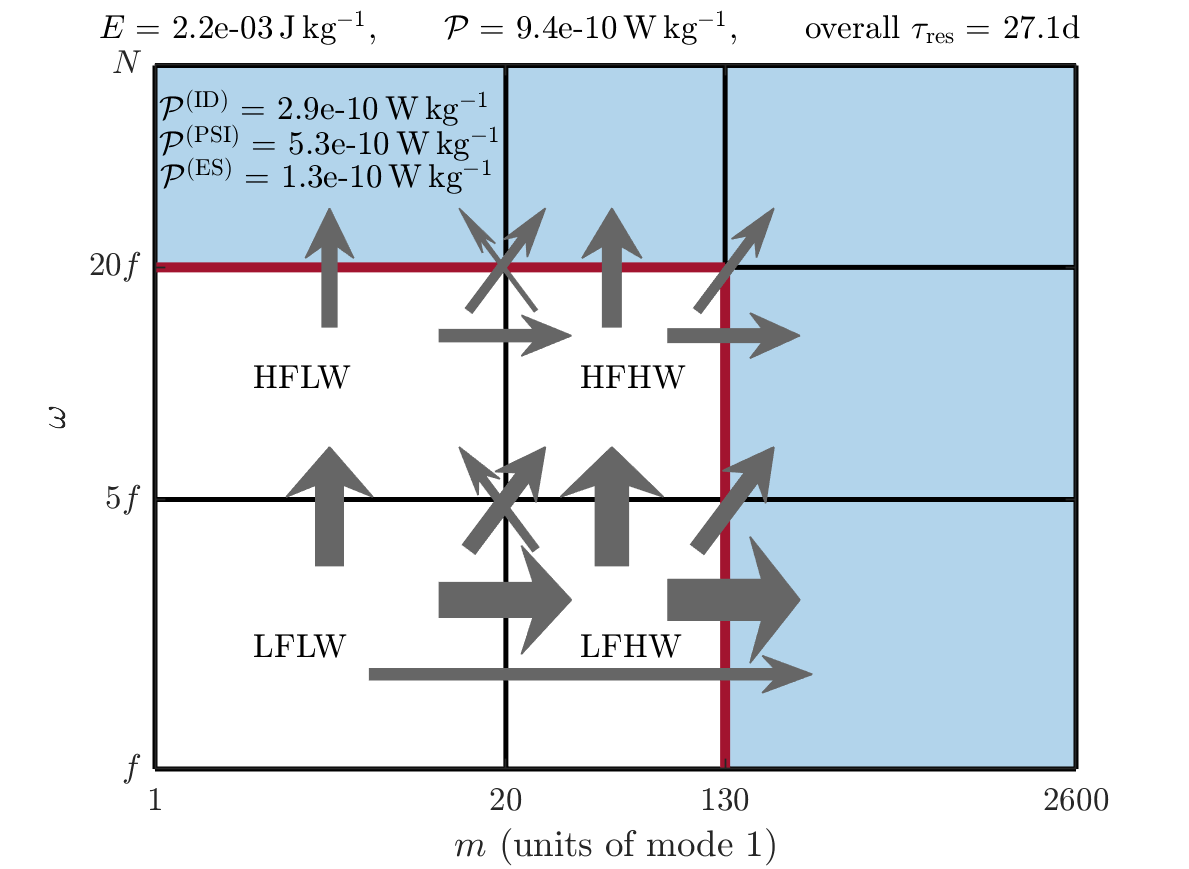}
\caption{Sensitivity test of evaluation of TKE production rate upon halving the value of the high-wavenumber boundary with the dissipation region. Both panels 
are evaluated for the same GM76 spectrum.}\label{fig:9M}
\end{figure*}

In light of our analysis of nonlinearity level, highlighting the presence of high nonlinearity at large vertical wavenumbers, it is important to study the sensitivity of our results with respect to the location of the boundary $m_c$ defining the dissipation region. If the flux across the boundary is insensitive to the position of the boundary itself, we are free to exclude the most nonlinear scales which hinder the weak-nonlinearity assumption. In Fig.~\ref{fig:9M} we repeat our analysis for the GM76 spectrum, by halving the value of $m_c$. We show that this changes the calculated value of TKE production rate by less than $5\%$, adding to the robustness of the theoretical calculation of $\mathcal P$. Also, notice that the increased nonlinearity at the transition to strong turbulence in proximity of $m_c$ is expected, since the instability mechanisms at the onset of turbulence itself are strongly nonlinear. It thus appears to be natural to have values of the nonlinearity parameter $r_{\rm nl}(m,\omega)$ of the order of unity in proximity of $m=m_c$.

{\color{black}In order to compute the transfers through the boundary $m_c$, we have to assign a spectral distribution in the dissipation region, past the $m_c$ boundary. Lacking knowledge, in general, of these scales, operationally we extend the spectral slope past the $m_c$ boundary. A key observation that justifies this operation resides in the fact that the transfers are dominated by local interactions, and we confidently estimate that most of the interactions transferring energy past $m_c$ involve wavenumbers that are smaller than $2m_c$. A sensitivity test performed in~\cite{regional} shows that cutting off completely all wavenumbers with $m>2m_c$ implies no perceived change in the time rate of change for wavenumbers with $m<m_c$. We do not repeat such a test here. We opted for this choice for simplicity, since the introduction of a sixth spectral parameter would make a full exploration of the spectral parameter space unattainable by our numerical method. Moreover, the poor observational handle on these small-scale wavenumbers would have made the uncertainty on this extra parameter very high.
On the other hand, the physicality of our choice relies on the fairly small bandwith in vertical wavenumber involved in the transfers past $m_c$, due to the local character of the wave-wave interactions.}

{\color{black} Similar in spirit, a sensitivity test on the location of the high-frequency threshold, which we placed at $20f$, was carried out in~\cite{dematteis2022origins}. They showed that moving the high-frequency boundary past which the energy transfers are computed by a factor of $\sqrt{2}$ implied a total difference in the turbulence production of no more than $10\%$. We refer to that test without repeating it here, since their non-rotating approximation is retrieved by our high-frequency regime, and our results generalize consistently the results of~\cite{dematteis2022origins}.}

{\color{black} We end this section by noting that the weak sensitivity to the location of the spectral boundaries in our framework has to be ascribed to the dominance of local interactions in determining the spectral transfers. This stands in contrast to the eikonal framework, in which large scale separation between small scale waves and large scale background shear implies a high sensitivity to the wavenumber thresholds of scale separation and of wave breaking~\cite{ijichi2017eikonal,polzin2022one}. The array of sensitivity tests here described ensures the robustness of our quantifications, as they do not depend upon arbitrary degrees of freedom that could serve as tunable parameters. Furthermore, they should not be substantially impacted by the gradual breakdown of weak wave turbulence near the spectral boundary $m_c$.}

\subsubsection*{Finestructure contamination}\label{sec:6.4.2}

The analysis of ocean data is predicated upon a Reynolds decomposition and vertical scale separation, for which we direct the reader to \cite{polzin2014finescale} for a discussion of  `best practices'.  The most problematic issue here is the assumption that the perturbation fields represent internal wave dynamics.  Experience from a variety of internal wave climates away from boundaries demonstrates that this assumption is more of an issue for finescale buoyancy gradients than for finescale shear \citep{polzin2003partition}, especially if the data contains healthy near-inertial signatures.  

Finescale buoyancy variability may have a quasi-permanent component associated with either (i) the generation of potential vorticity anomalies at boundaries by topographic torques and the consequent injection of those anomalies into the ocean interior \citep{kunze1993submeoscale}, (ii) generic upper ocean variability, either resulting from patchiness in forcing and restratification \cite{joyce1998meso} or through the development of submesoscale variability \cite{thompson2016open}, (iii) internal wave-driven scarring of the thermocline on vertical scales larger than those of overturning events \citep[][section 5.2]{polzin2004isopycnal}, and (iv) up-gradient buoyancy fluxes associated with double-diffusive phenomena that spontaneously creates finescale variability \citep{stern1960salt, schmitt1994double}.  In (i)-(iii) the dynamics of quasi-permanent finestructure is that of either stratified or rotating stratified turbulence, in which a horizontal wavenumber cascade is attained via horizontal stresses and horizontal rate of strain, with vertical coupling at scales limited by shear instabilities \citep{riley2000fluid, kunze2019unified}.  

\begin{figure}[h]%
\centering
\includegraphics[width=0.8\linewidth]{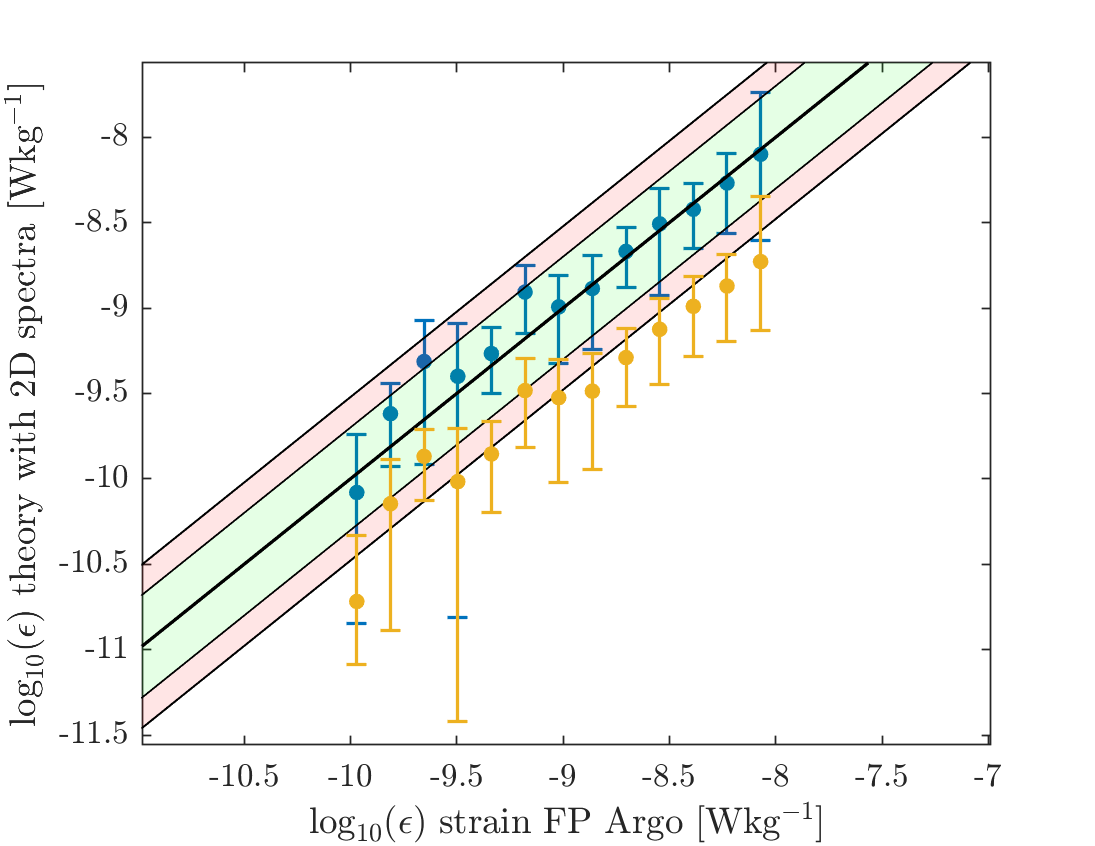}
\caption{Sensitivity test upon adding an extra $0.20$ to the vertical wavenumber slopes $s_m$ (yellow dots) obtained from the Argo strain spectra, compared to the results in Fig.~3{\bf B} (blue dots), in an attempt to roughly bound the effects of finestructure contamination. }\label{fig:10M}
\end{figure}

The coupling of internal waves with quasi-permanent finestructure through a wave breaking process (issue (iii)), however, implies our vertical wavenumber power law estimates are likely to be consistently biased high in association with an increasing contribution of quasi-permanent finestructure with increasing wavenumber.  There is one reference analysis to constrain this potential bias:  this comes from analysis of High Resolution Profiler data in the North Atlantic tracer Release Experiment (NATRE).  In this instance, the observed vertical wavenumber spectral slope for velocity is -2.55.  After subtracting the quasi-permanent signatures, the exponent is -2.75.  

We use this difference as a metric of potential bias, with the insight that the spectral slopes for the NATRE data set are an outlier in vertical wavenumber - horizontal wavenumber space \cite{regional} and that such bias cannot be properly assessed without a data set that combines high quality velocity and density finestructure.  The result of the analysis upon adding the potential bias of 0.20 to the vertical wavenumber slopes from Argo is shown in Extended Data Fig.~\ref{fig:10M}. 
These new estimates of $\epsilon$ are generally lower than the original ones, roughly by a factor 3-4, which also impairs the agreement with the reference data set \cite[horizontal axis;][]{pollmann2023}. Note, however, that this reference data set is also derived from Argo float observations, assuming a constant wavenumber slope $s_M$ of 2. In a fully consistent comparison, 
we would have needed to account for the wavenumber slope bias also in this reference, which we do not attempt here owing to the high costs of creating such a global data set.

Regarding issue (iv), we follow \cite{pollmann2020global} and investigate the sensitivity of the wavenumber slope estimates to the Turner angle \cite{ruddick1983practical}. The Turner angle Tu is defined as

\begin{equation}
    Tu = atan \left(\alpha \frac{\partial \Theta}{\partial z} - \beta \frac{\partial S_A}{\partial_z}, \alpha \frac{\partial \Theta}{\partial z} + \beta \frac{\partial S_A}{\partial_z} \right),
\end{equation}
where $\Theta$ is conservative temperature, $S_A$ absolute salinity, $\alpha$ the thermal expansion coefficient and $\beta$ the haline contraction coefficient. Double-diffusive convection is characterized by Turner angles between 45$^{\circ}$ and 90$^{\circ}$, where positive angles represent the saltfinger and negative angles the diffusive regime. Both regimes describe an instability emerging in hydrostatically stable density profiles due to the vastly different molecular diffusivities of the tracers determining the density \citep[i.e. heat and salt in the ocean, see e.g.][]{stern1960salt}. 

\begin{figure}[h]%
\centering
\includegraphics[width=\linewidth]{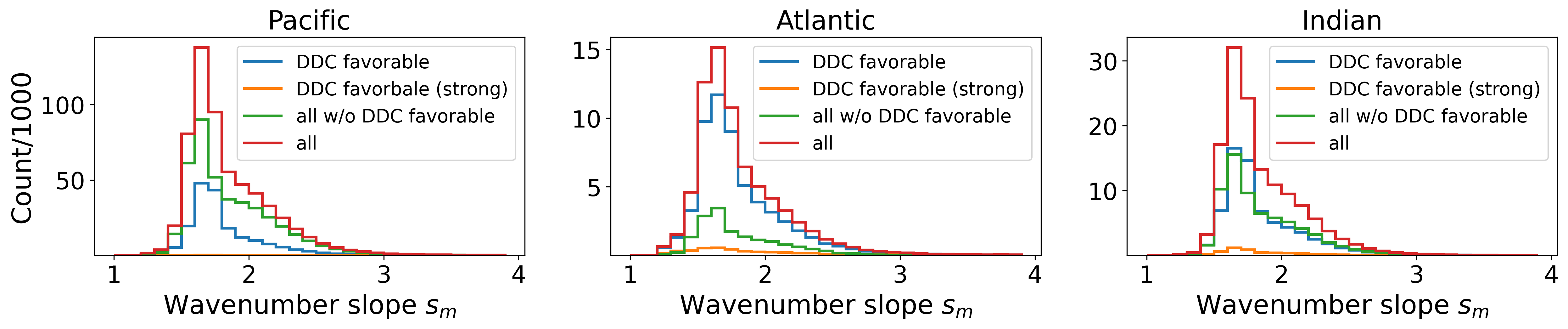}
\caption{Distribution of wavenumber spectral slope $s_M$ in the Pacific, Atlantic and Indian Oceans, using (a) profiles with at least 50\% of the Turner angle estimates meeting $45<\|Tu\|<90$ (DDC favorable); (b) profiles with at least 50\% of the Turner angle estimates meeting $75<\|Tu\|<90$ (favorable to strong DDC); (c) all profiles except those identified in (a); (d) all profiles. }\label{fig:11Tu}
\end{figure}
As illustrated in fig.\,\ref{fig:11Tu}, there is no clear change in the histograms when wavenumber slopes are estimated from profiles favorable to DDC compared to using all profiles or only those without a DDC imprint. We thus conclude that finescale contamination by DDC (issue iv) is not an important source of uncertainty for the wavenumber slope estimates, which represent averages over more than 2 million individual estimates.

\pagebreak



%\bibliography{references}% common bib file


\section*{Acknowledgments}

Discussions with D. Olbers, S. Purkey, A. Thurnherr, Y. C. Wu, Y. Pan,  and J. Skitka are gratefully acknowledged. %We are grateful to Yue Cynthia Wu and two other anonymous reviewers for their thorough and constructive comments that helped us to improve the manuscript. 
We acknowledge the International Argo Program and the national programs that contribute to it.  (https://argo.ucsd.edu,  https://www.ocean-ops.org). We would like to acknowledge the scientists and technicians preparing, deploying,
and recovering the moorings, whose data are collected in the GMACMD database. We would also like to acknowledge
the captains and crews of the many research vessels supporting
all of these missions. GD and YVL gratefully acknowledge funding from NSF DMS award 2009418. GD and KP gratefully acknowledge funding from grant OCE-83243900. {\color{black}GD acknowledges funding from the Simons Collaboration on Wave Turbulence.} \textcolor{black}{FP acknowledges the Collaborative Research Centre TRR181 ‘Energy Transfers in Atmosphere and Ocean’ funded by the Deutsche
Forschungsgemeinschaft (DFG, German Research
Foundation)—Projektnummer 274762653.} \textcolor{black}{CW was funded by OCE-2048660 and ONR-N00014-18-1-2598.}

\section*{Authors contributions statement} GD performed the theoretical/numerical calculations with support from YL. ALB, FP, and GD developed the data analysis to build the global combined dataset of 2D internal wave spectra. GD and KP performed the HRP data analysis. GD, FP, and ALB wrote the first draft of the manuscript. GD, ALB, FP, KP, MA, CW, and YL  contributed to designing the research, drafting sections, and revising the manuscript.

\section*{Competing interests statement} The authors declare no competing interests.

\end{document}